\documentclass[useAMS,usenatbib]{mn2e}
\usepackage{aas_macros}
\usepackage[a4paper,centering, totalwidth=520pt, totalheight=700pt]{geometry}
\def\k#1 {k_{{\rm #1}}}

\def\mH2p{H_2^+}

\def\ltsima{$\; \buildrel < \over \sim \;$}
\def\simlt{\lower.5ex\hbox{\ltsima}}   
\def\gtsima{$\; \buildrel > \over \sim \;$}
\def\gtsim{\lower.5ex\hbox{\gtsima}}

\usepackage{graphicx}
\usepackage{amssymb}
\usepackage{amsmath}

\title[Tracing the dark matter sheet]{Tracing the Dark Matter Sheet in Phase-Space}
\author[T. Abel, O. Hahn \& R. Kaehler]{Tom Abel\thanks{Email:
    tabel@stanford.edu}, Oliver Hahn\thanks{Email:
    ohahn@stanford.edu} \& Ralf Kaehler\thanks{Email:
    kaehler@slac.stanford.edu}$^1$\\
$^{1}$Kavli Institute for Particle Astrophysics and Cosmology,
Stanford University, \\
SLAC National Accelerator Laboratory, Menlo Park, CA 94025}

\date{MNRAS submitted}

\begin{document}
\pagerange{\pageref{firstpage}--\pageref{lastpage}} \pubyear{2011}
\maketitle

\label{firstpage}

\begin{abstract}
  The primordial  velocity dispersion of dark matter  is small compared to
  the velocities attained during structure formation. The initial
  density distribution is close to uniform and it occupies an initial
  sheet in phase-space that is single valued in velocity
  space. Because of gravitational forces, this three dimensional
  manifold evolves in phase-space without ever tearing, conserving
  phase-space volume and preserving the connectivity of nearby
  points. N--body simulations already follow the motion of this sheet
  in phase-space. This fact can be used to extract full fine-grained
  phase-space structure information from existing cosmological N--body
  simulations. Particles are considered as the vertices of an
  unstructured three dimensional mesh moving in six dimensional
  phase-space. On this mesh, mass density and momentum are uniquely
  defined. We show how to obtain the space density of the fluid,
  detect caustics, and count the number of streams as well as their
  individual contributions to any point in configuration-space. We
  calculate the bulk velocity, local velocity dispersions, and
  densities from the sheet -- all without averaging over control
  volumes. This gives a wealth of new information about dark matter
  fluid flow which had previously been thought of as inaccessible to
  N--body simulations. We outline how this mapping may be used to
  create new accurate collisionless fluid simulation codes that may be
  able to overcome the sparse sampling and unphysical two-body effects
  that plague current N--body techniques.
\end{abstract}


\maketitle

\section{Motivation}

For the past 40 years, N--body simulations have allowed to numerically
study the evolution of the distribution of matter in the expanding
Universe \citep[cf.][]{Peebles:1971a, Bertschinger:1998,
  Springel:2005a}. A significant number of simulation codes have been
developed for this purpose \citep[e.g.][to name just a
few]{Efstathiou:1985, Couchman:1991, Bryan:1997, Stadel:2001,
  Springel:2001b, Teyssier:2002, Wadsley:2004}. All such approaches to
structure formation model the collisionless fluid of dark matter by a
set of massive particles (typically of equal mass) and differ in how
the gravitational forces are calculated at the positions of the
particles. The forces are applied to update the velocities which in
turn are used to update the positions. The system is then evolved
forward in time. From such simulations much has been learned about the
formation and evolution of cosmological structures and they have
become a standard tool in physical cosmology.  While three dimensional
calculations have difficulty in sampling the six dimensional
phase-space well \citep[see e.g.][]{Buchert:1991} they have found a
very large range of applications and have driven much of the progress
that has been made in the past decades of understanding structure
formation. In quite a range of these applications the space density of
the dark matter fluid is required and in many others the phase-space
density is of great importance.

Some open questions that require detailed information about the dark
matter density and its velocity distribution are related to dark
matter detection. For the indirect detection techniques, the
predictions of the dark matter annihilation luminosity depend
sensitively on the density of the dark matter streams and the
distribution and the relative velocities of the particles. Assuming
well mixed phase-space and assuming a shape of the velocity
distribution function, this annihilation rate would scale with the
square of the space density, $\rho$.  In current work, these estimates
are typically carried out by fitting spherical profiles to the main
dark matter halo and its subhaloes and then assign annihilation
luminosities by scaling the square of the smoothed halo and subhalo
profiles appropriately. This smoothes the dark matter fluid
sufficiently to avoid noisy estimates of the annihilation signal
\citep{2008Natur.454..735D, 2008Natur.456...73S}. However, in general
the annihilation rates depend on the relative velocity of the dark
matter particles interacting. Here one can distinguish between dark
matter annihilation within individual streams of dark matter as well
the contribution from stream-stream interactions which differ strongly
in the relative velocities \citep[see e.g.][]{Hogan:2001,
  Afshordi:2009}.  This fine grained dark matter phase-space structure
is equally important for considerations of dark matter direct
detection experiments where one applies velocity cuts in the
experimental analysis in order to reject certain backgrounds. The
annular modulation of the relative velocity and the main dark matter
streams in the solar neighborhood provided by the earths motion around
the sun allows one to potentially map the fine grained phase-space
structure should the experiments be able to detect dark matter.

The best current approaches to probe phase-space structures were
surveyed by \cite{2009MNRAS.393..703M}. These typically start with some
tessellation of phase-space such as a Delaunay triangulation or a
Voronoi tesselation \citep{1996MNRAS.279..693B}, or cartesian trees
\citep{2006MNRAS.373.1293S, 2010CoPhC.181.1438A}. The mass of the
particles found within the cells give the local densities.  When only
the space density is required, adaptive kernel smoothing is often
employed.  In fact, most images of dark matter simulations shown are
projections of kernel smoothed particle distributions. Configuration
space density estimators also play a particularly important role when
studying the topology and character of the cosmic web
\citep[e.g][]{Schaap:2000,Pelupessy:2003, Aragon-Calvo:2007,
  Colberg:2008,Neyrinck:2008}. However, in all cases, control volumes
are defined such that they contain sufficient numbers of particles to
reduce sampling noise. Unfortunately, this will average over large
regions of configuration and phase-space and consequently effectively
degrade the spatial resolution of the calculation.

In any case, there is ample motivation to study the distribution,
evolution and current state of dark matter in the Universe further by
observations as well as simulations. In this contribution, we introduce
a novel way to analyze $N$-body simulations. Our approach naturally
arises from considering how the collisionless fluid of dark matter is
expected to evolve in phase-space.

As is well known, \citep[e.g.][]{Shandarin:1989}, at early enough times,
i.e. before shell crossing, the motion of the dark matter fluid is well
described by the Zel'dovich approximation \citep{ZelDovich:1970} as a
potential flow
\begin{eqnarray}
\mathbf{x}_t & = & \mathbf{q} + g_t\,\boldsymbol{\nabla}\phi(\mathbf{q}),\\
\mathbf{v}_t & = &\dot{g}_t\,\boldsymbol{\nabla}\phi(\mathbf{q}).
\end{eqnarray}
Here $\phi$ is a potential field that is proportional to the initial gravitational
potential field of perturbations, $\mathbf{q}$ are the initial particle positions and $g(t)$ is
the growth factor of linear perturbations. At early times, i.e. for $t\to 0$,
also $g\to 0$. These particles occupy a three dimensional submanifold $S$
of the entire six dimensional phase-space with the time-dependent mapping:
\begin{equation}
\mathbf{q} \mapsto \left(\mathbf{q}+ g_t\,\boldsymbol{\nabla}\phi(\mathbf{q}), \dot{g}_t\,\boldsymbol{\nabla}\phi(\mathbf{q})\right),
\end{equation}
where for $t\to0$
\begin{equation}
\mathbf{q} \mapsto \left(\mathbf{q}, \dot{g}_0\,\boldsymbol{\nabla}\phi(\mathbf{q})\right),
\end{equation}
the three-dimensional structure ($\mathbf{q}\in\mathbb{R}^3$) can be easily seen.  
The map between $\mathbf{x}_t$ and $\mathbf{q}$ is
bijective until shell crossing occurs, i.e when more than one stream
of dark matter exists at one spatial location.  We will refer to this
three dimensional submanifold as the dark matter sheet (even when
discussing it in one or two spatial dimensions). The volume of the
sheet continues to grow as structure forms and
evolves \citep{Shandarin:1989,Vogelsberger:2008}.

At the same time, current $N$--body simulations of structure formation
do already follow individual dark matter particles through phase-space
\citep[e.g.][]{Bertschinger:1998}. The $N$-body technique thus
corresponds to sampling the sheet at a finite number of points
$\mathbf{q}$ with the entire mass concentrated at their positions.

\cite{Vogelsberger:2008}, \cite{White:2009} and \cite{Vogelsberger:2011} developed a
powerful approach to augment cosmological simulations to record more
knowledge about the evolution of this dark matter sheet.  They derive
an equation of motion for the distortion tensor around every particle
to linear order and then evolve it with every particle during the
simulation. They refer to this as the geodesic deviation equation
(GDE). This gives access to information about the evolution
of the stream density along every particle trajectory and allows to
track the number of caustics an infinitesimal fluid element surrounding
a dark matter particle will experience. This technique goes a long way
in obtaining more information about the fine grained phase-space
structure in dark matter haloes \citep{Vogelsberger:2011}.

More restricted calculations in this context have been carried out in
fixed potentials \citep{Stiff:2001} or one dimensions
for Newtonian gravity \citep{Alard:2005} and General Relativity
\citep{Rasio:1989}. Also in the context of stellar dynamics there is a
large body of literature which  explores  details of the
phase-space structure of stellar system. From a numerical point of
view the work of \cite{Cuperman:1971a} is particularly
remarkable. Some 40 years ago these authors realized that in one
dimension one can follow the phase-space boundary of a collisionless
fluid and they give a beautiful implementation and calculations 
treating the phase-space fluid as a continuum. Studying the connection
of their formalism to the N--body technique is revealing and what
follows here is in some ways the extension to three dimensions with
the exception of our approach to velocity dimensions and the way the
Poisson equation is solved. 

We suggest that the three-dimensional manifold can be decomposed by a
space-filling grid that connects a finite number of vertices
$\mathbf{q}$. The simplest version is to decompose the volume into
three dimensional simplices, i.e. tetrahedra, which have the nice
topological property of being either convex or degenerate. For any
choice of a regular lattice of vertices $\mathbf{q}$, such a
tetrahedral decomposition can be achieved by a Delaunay triangulation.
In contrast to the particle discretization, we can now think of the
dark matter mass being spread out over the corresponding volume
elements.

This mesh traces the dark matter sheet as it subsequently evolves in
phase-space. The motion of the mesh vertices are evolved using the
Vlasov-Poisson equation of motion leading to complex foldings of the
submanifold \citep[e.g.][]{Arnold:1982, Tremaine:1999}. Any such
folding is coincidental with a volume inversion of a simplex.  This
volume inversion occurs when the simplex topologically evolves through
a degenerate state (where the tetrahedron is planar because one vertex
moves through the plane defined by the remaining three) which is
equivalent to the emergence of a caustic. Note how this corresponds
exactly to the sign changes of the distortion tensors of
\cite{Vogelsberger:2008} (their eq. 24).

The motion of the vertices does not change the connectivity of the
mesh so that at all times the simplex structure can be constructed
from knowledge of the $\mathbf{q}$. For cosmological $N$-body
simulations, there exists a unique mapping between a particle ID and
$\mathbf{q}$, so that the phase-space structure of the dark matter
sheet can be reconstructed at all times. Projecting the sheet onto
configuration space gives then a volume filling density field of the
dark matter fluid that we propose to use as the density field that
should be used to solve Poisson's equation in future solvers for
collisionless fluids. Current $N$-body solvers do not evolve the
vertices consistently with a density field construed in the proposed
way.

As a first step towards this goal, we analyze the results of standard
cosmological $N$-body simulations using this new definition of the
dark matter sheet. The plan of the paper is as follows. First we will
explain one and two dimensional analogues to introduce the relevant
concepts. We then describe the details of our implementation before we
apply the method to analyze cosmological large-scale structure as well as
the phase-space properties of a single galaxy cluster halo. 

\section{Evolution of three-dimensional sheets in phase-space}

The distribution function $f(\mathbf{x}, \mathbf{p}; t)$ describes the density
of a fluid in phase-space. It evolves via
\begin{equation}
\frac{\partial f}{\partial t} = - \frac{\mathbf{p}}{m} \cdot \boldsymbol{\nabla}_x f -
\boldsymbol{\nabla}_x \phi \cdot \boldsymbol{\nabla}_p f, 
\end{equation}
where $\phi$ is the gravitational potential and $m$ is the dark matter
particle mass. Fluid elements get stretched or compressed in coordinate space by
advection $\frac{\mathbf{p}}{m} \cdot \nabla_x f$, and in the momentum
coordinates by the gravitational forces $\frac{\mathbf{p}}{m} \cdot
\boldsymbol{\nabla}_x f$. Note that in a Lagrangian frame, the first term on the
right hand side is zero. Furthermore, the second term describes how
the fluid is stretched in momentum space and does not affect the space
density of the fluid parcel. This just states Liouville's theorem
\citep{gibbs1902elementary} that the volume in phase-space is
conserved. Hence, any fluid volume $\triangle \mathbf{x} \triangle
\mathbf{v}$ will remain constant. We are interested here in the space
density of the fluid, the projection of $f$ into coordinate
space. i.e. the integral $\rho(\mathbf{x}) = \int f(\mathbf{x}, \mathbf{v})
{\rm d}^{3}v$. The contribution to the space density of any stream of dark
matter is only affected by the volume it occupies in the space
coordinates, i.e. $\triangle \mathbf{x}$. Consequently, all that is necessary
to follow the evolution of the dark matter density is to follow the
Lagrangian evolution of fluid elements. The mass inside a volume
element is conserved and its contribution to the space density of dark
matter is described by the volume it occupies. Conversely, for a given
WIMP model one knows the initial velocity dispersion at any point in
space \citep[e.g.][]{Hogan:2001, Vogelsberger:2008}. Therefore, if one
knows the spatial part of the phase-space density one has information
about the density in velocity space. For a given shape of the initial
distribution function in the velocity directions (e.g. a Maxwellian)
one has a reliable measure of the intrinsic velocity density at all
times.

Using the Vlasov equation to describe DM is justified for most
particle physics inspired models of dark matter. For a WIMP scenario
with a $100\,$GeV particle e.g. there would be $10^{67}$ such
particles in the Milky Way alone. So an element in phase space that
contains a million such WIMPS, giving a well defined phase space
density, would have a spatial extent at mean density equivalent to a
cube 500 meters on a side. Such a volume would only be a few meters on
a side at the DM density expected in the solar neighbourhood. Clearly
we are interested in scales much larger than this and the
approximation of using a density in phase space is justified to a high
degree of confidence.

It is instructive to first describe a straightforward and well known
one--dimensional example of the evolution of a collisionless fluid
from which a number of lessons can be learned which apply equally well
in higher dimensions.

\subsection{The Zel'dovich pancake}
\begin{figure}
\centerline{\includegraphics[width=0.47\textwidth]{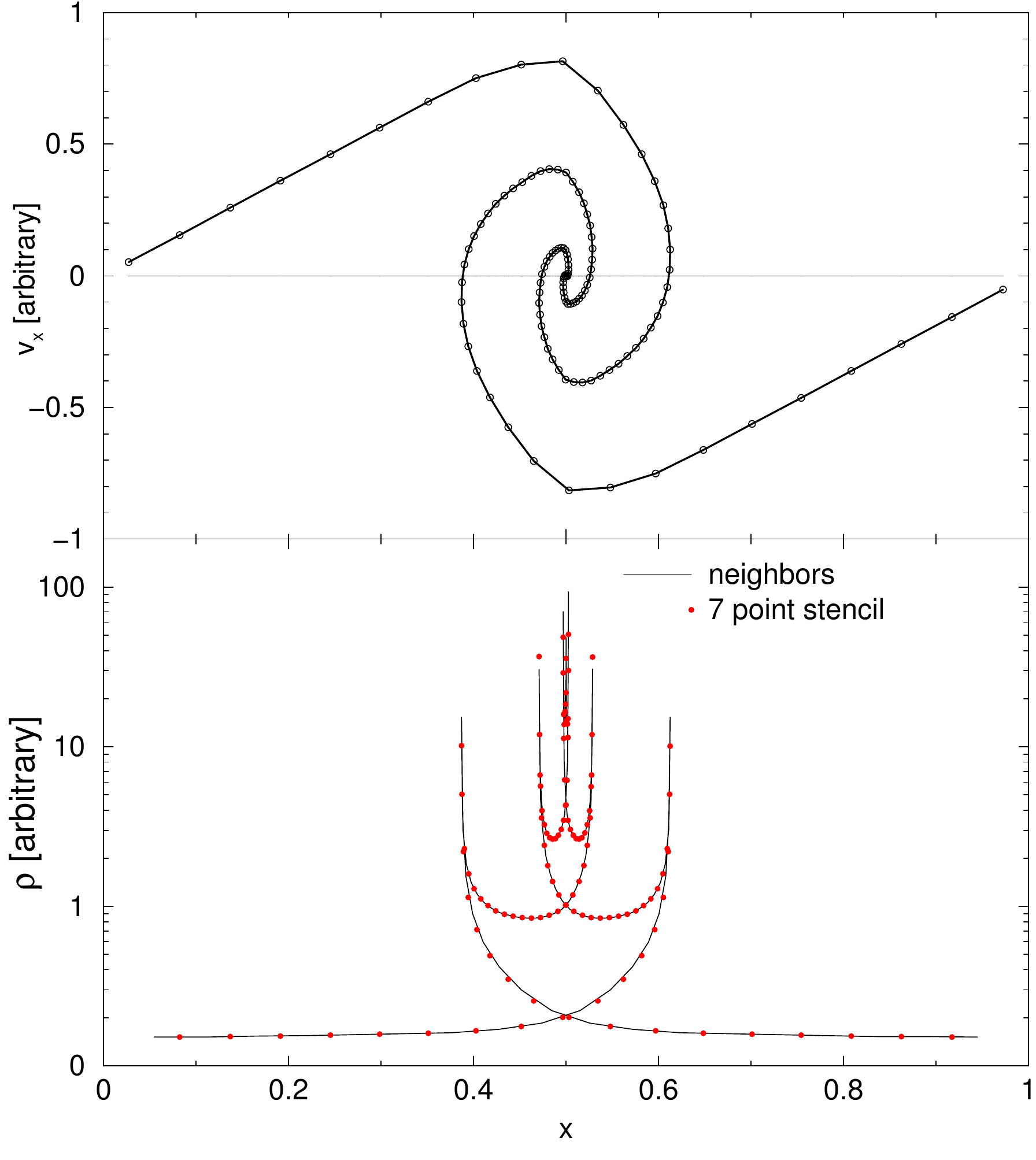}}
\caption{The one--dimensional plane wave collapse of Zel'dovich
  \citep{ZelDovich:1970, Binney:2004}. The top panel gives the phase-space diagram showing the velocities of the particles at their
  locations. The bottom panel gives the density of the dark matter
  inside the stream, one computed with a seven point stencil (red
  squares), and the other computed from the volume between two neighboring
  points (solid line). Knowing the spatial volume between particles
  along one stream is sufficient to obtain accurate density estimates
  at and between the points.  }\label{fig:zeldo}
\end{figure}

The phase-space diagram and the evolved density in a Zel'dovich plane
wave collapse is shown in Figure~\ref{fig:zeldo}. The initial sheet at
very early times would be coincident with the $x$-axis as the initial
velocity perturbation is small and the initial state models a nearly
homogeneous Universe. Sampling this initial state with particles of
equal mass results in a grid of uniformly placed particles. Their
configuration space volume is now simply related to their distances in the
$x$--direction. Figure~\ref{fig:zeldo} shows the results of computing
their local stream density from two approaches. In one, labelled
``neighbour'', we take the $V_i=x_{i+1}-x_i$ as the volume between
particle $i$ and $i+1$. One full particle mass is distributed in this
volume and the density at $(x_{i}+x_{i+1})/2$ is given by
$\rho_{neigh} = m_p/|V|$. The values shown as ``squares'' in the same
figure are computed including information from points further along
the stream, $\rho_{7pt}=6\,m_p/|x_{i+3}-x_{i-3} |$. It is defined at
the particle position $x_i$. A number of observations can be
made. Volumes defined in this way may be positive or negative
depending on whether particles have the same or opposite ordering that
they had initially. Volume elements may also become zero.  The density
involving more points along the stream gives rise to some smoothing
and density extrema are clipped. The central high configuration space
densities are reached for two reasons. The primordial stream densities
along the sheet become larger and many streams overlap adding their
densities. The number of streams in space is always an odd number at
any location in space. Only at the caustics may one measure even
numbers.

\begin{figure}
\centerline{\includegraphics[width=0.37\textwidth]{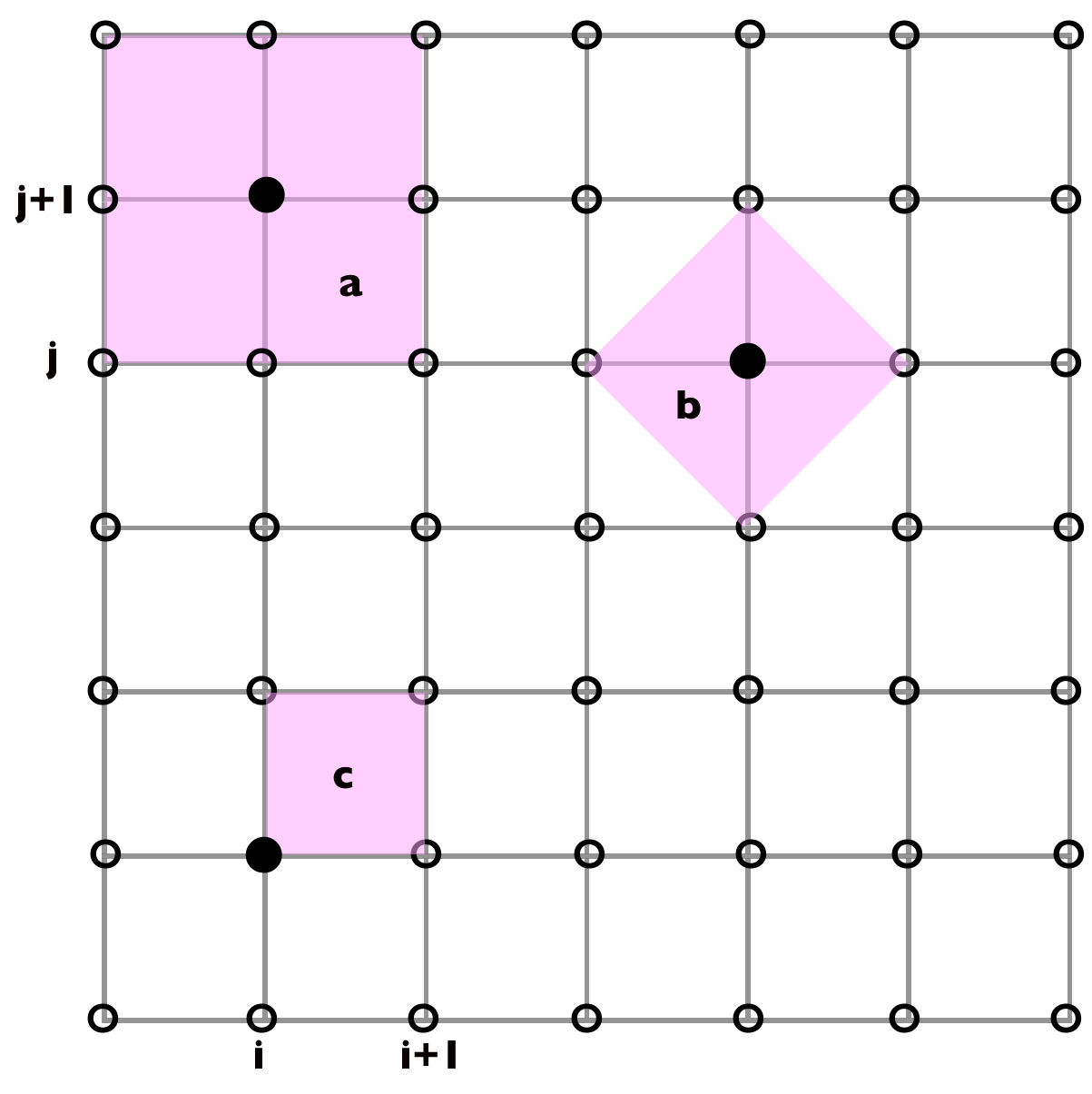}}
\caption{There are a number of possible cell elements (and that can be
further decomposed into simplices) one may consider as the
  unit cell. These unit cells can be used to tessellate the three dimensional dark
  matter sheet which connects the particles of an N--body simulation.
   }\label{fig:choices}
\end{figure}

The particle locations trace the sheet in phase-space. Any
unstructured space-filling grid that connects adjacent fluid elements
may be used to trace the dark matter sheet as it evolves in
phase-space. In fact, there is significant ambiguity here as
illustrated in Figure~\ref{fig:choices}. The two--dimensional analog
shown there has is based on triangles (the 2D simplex). The smallest possible elements
one may thus choose to follow would be the Delaunay triangulation of the
points. However, these would give two resolution elements per square
initial cell (case c). It would seem unreasonable that the mass of the fluid
would be conserved exactly in each element, given that we only have
information at the vertices. This could only be true if the mesh
points were not distorted very much and the gradients of the flow,
both in configuration and momentum space had length scales much larger
than the sides of the triangle. However, cases a) depicted in the
figure would likely be a better choice as a fundamental resolution
element as its eight nodes on the surface would be able to more
accurately describe the deformations caused by the flow pattern. The
mass inside that boundary would be conserved to a better degree than
choices for a fundamental resolution element with smaller area. Note,
however, that we still can use triangles to calculate the volume (area
in 2D) of the sheet.

After considering some preliminaries in one and two dimensions, let us
proceed to the three--dimensional case.


\subsection{An Unstructured Grid to trace the Dark Matter Sheet in
  phase-space}

\begin{figure}
\centerline{\includegraphics[width=0.37\textwidth]{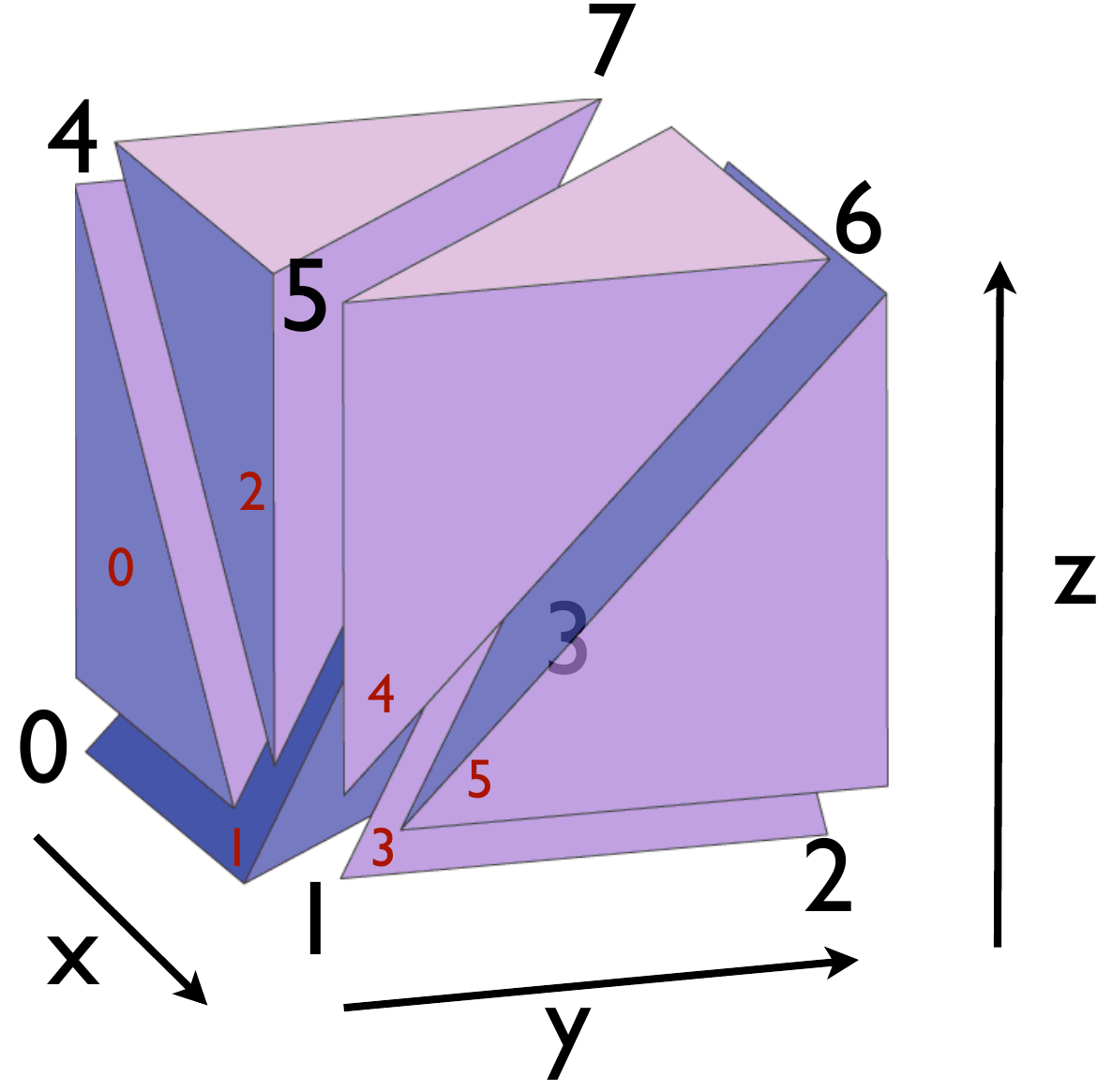}}
\caption{One of the possible decompositions of a cubical cell into
  tetrahedra. This choice results in 6 tetrahedra all of equal volume
  fraction. This particular case has two of the cube corners as vertices for
  all tetrahedra and while the other corners are connected to two
 tetrahedra each.  }\label{fig:tets}
\end{figure}

The simplex in three dimensions is the tetrahedron. We will use it to
calculate volumes and tesselate our chosen fundamental volume
element. This is exactly equivalent to choosing line elements in one
and triangles in two dimensions as the elements which are summed in
volume calculations. Figure~\ref{fig:tets} shows one of the choices we
have employed to tesselate a cubical fundamental cell. The figure also
gives a numbering of vertices of the six tetrahedra which make up the
cell. The connectivity of vertices is chosen such that in a regular
uniform grid all tetrahedron volumes are positive. For much of the
calculation we keep the sign, because, as we will see, it can be a
useful diagnostic of the flow. If we shift a tetrahedron such that one
vertex coincides with the origin, the volume is simply given by the
determinant or, equivalently, from a scalar and a cross product
involving its other three vertices, $V_{tet}=\det\left|
  \mathbf{a},\mathbf{b},\mathbf{c}\right|/6=\mathbf{a}\cdot(\mathbf{b}
\times \mathbf{c})/6$.  This implies that the volume can be negative
if the tetrahedron has been turned inside out. This sign inversion
occurs when one vertex moves through the plane defined by the other
three vertices of the tetrahedron and is an efficient way to find
caustics. Within an N-body calculation, tracking the number of volume
inversions could be used to trace the
number of caustics crossed by a fluid element. This volume can now
be used straightforwardly  to estimate the local stream density of
the fluid element described by the tetrahedron. In our case, one tetrahedron
contains one sixth of the mass of an N-body particle spread over
its volume so that
\begin{eqnarray}
\rho_{s} = \frac{m_P/6}{V_{tet}} =  \frac{m_P}{|\mathbf{a}\cdot(\mathbf{b} \times \mathbf{c})|}.\label{equ:rhostream}
\end{eqnarray}
Alternatively one may choose a cubical region rather than a single tetrahedron
as the fundamental volume element to consider. This can be achieved e.g. by using the 24
($2\times 6+6\times2$, see Figure~\ref{fig:tets})
tetrahedra around each point which abut to a given vertex. Their
volumes are then thought of containing four times the mass of one
particle. This effectively averages the density field on a kernel of
the same size as four cubical volumes.

\begin{figure}
\centerline{\includegraphics[width=0.47\textwidth]{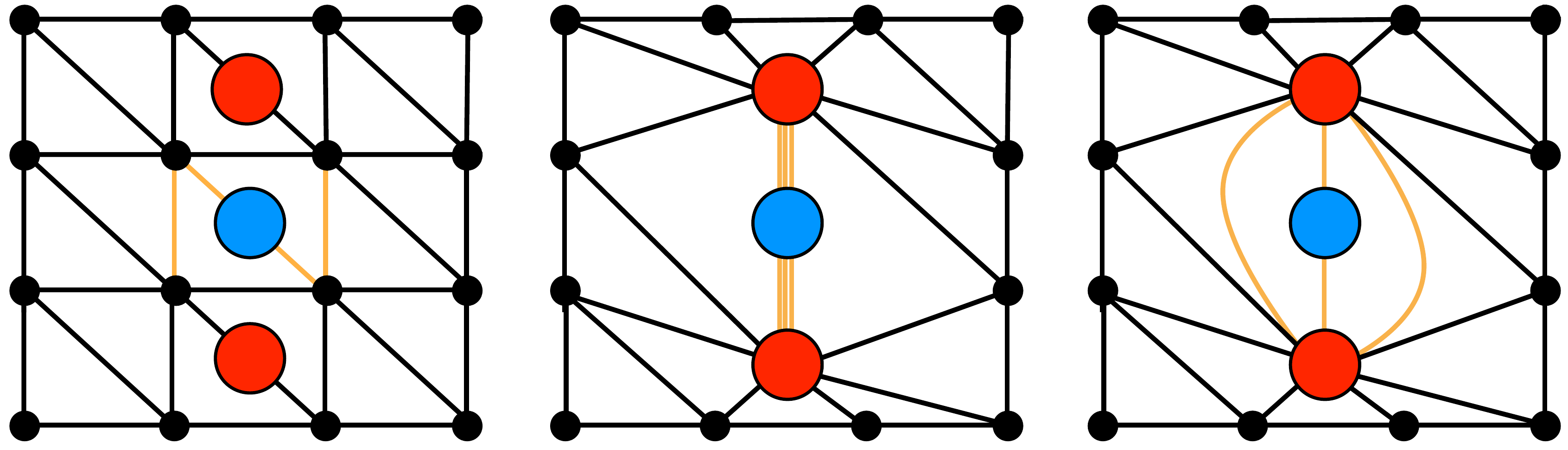}}
\caption{The need to resolve critical points of the flow: (left) a
  halo-void-halo configuration that, when probed with a tetrahedral
  structure leads to a final state of the structure as given in the
  middle panel. The correct motion of the fluid element could have
  been more correctly been described by the right panel.  The loss of
  spatial resolution can be circumvented by smoothing or by adaptive
  refinement of the volume elements.  }\label{fig:saddlepoints}
\end{figure}

In a simple implementation, the vertices of the tesselation correspond
to the particles of a standard $N$-body simulation. This implies that
the vertices are moving in a Lagrangian way so that the spatial
sampling is degrading over time in low-density regions and improving
in high-density regions.  One implication of the Lagrangian motion of
the mesh vertices is that unresolved volume elements may not evolve
according to the properties of the underlying flow. One particular
such example is given in Figure~\ref{fig:saddlepoints} which
illustrates that volume elements covering a divergent flow can evolve
as if they were in a convergent flow if only the vertices are convergent.
This behaviour implies that single volume elements, i.e. tetrahedra, of
our space discretization  are not to be trusted as perfect bags
of fluid. In the example from Figure~\ref{fig:saddlepoints}, the
single tetrahedron suggests that shell-crossing occurred across a
divergent flow which is unphysical -- e.g. implying spurious links
between haloes across unresolved void regions in cosmological context.

There are two possibilities to achieve a more robust density estimate:
(1) The volume estimates of several neighbouring tetrahedra are combined, so that
density estimates are based on a larger region of the flow, or (2)
local refinement of the volume elements is performed whenever e.g.
axis ratios or curvature in phase-space suggest that the volume
element might miss the flow properties.  The second option is certainly
the more exciting possibility to achieve a density estimation method
that is well consistent with flow properties. It requires however that
the discretized dark matter sheet is refined while the $N$-body
simulation is run, so that the newly inserted vertices are evolved
with the flow. For this reason, we focus on an evaluation of the
averaging approach in this paper and consider refinement strategies in
a future publication.

\subsection{Implementation: Computing the stream properties}

The choice of tetrahedra used for most of the plots in this paper were
given with a slightly different tesselation than the one shown in
Figure~\ref{fig:tets}. The connectivity list for the six tetrahedra
specifically is [4, 0, 3, 1], [7, 4, 3, 1], [ 7, 5, 4, 1], [7, 2, 5,
1], [7, 3, 2, 1], [7, 6, 5, 2] where the unit cube vertices are
labeled as in Figure~\ref{fig:tets}. A natural way to obtain the dark
matter stream density at the location of an N-body particle is to
consider the volume surrounding any vertex $\mathbf{x}$ as the union
of all the tetrahedra that share this vertex. The volume of this union
is computed using the modulus of the volumes, i.e. inverted tetrahedra
are not subtracted. Such a volume carries the mass of four particles
which allows us to estimate the local density at a particle,
$\rho_{s,p} = 4\, m_p/\sum_{i=1}^{24}V_{tet}^i$.  We call this local average over
adjacent tetrahedra the ``primordial'' stream density at a
particle's position. It is a well defined quantity even after
shell-crossing. However, it is only defined for the vertices
$\mathbf{x}$ of the tessellation. The simpler definition of
equation~(\ref{equ:rhostream}), however, works very well and, from 
visual inspections of dark matter renderings, appears perfectly justified. 

The next step in calculating a configuration space density estimate,
as well as in evaluating other stream properties is an integration
through velocity space. This is achieved by finding all intersections
of tetrahedra with the point $\mathbf{y}$ at which the density or
other properties are to be determined. We refer to all these
intersections, which are not part of the primordial (or fundamental)
stream, as the ``secondary'' streams.  Because we start from a complete
tessellation of the sheet, the number of tetrahedra enclosing any
spatial point is the number of streams contributing at this point.

At the heart of a fast algorithm is thus a way to speed up this search
for point-tetrahedron-collisions. To find the phase space properties
e.g. at the locations of all the simulation particles one at first
sight expects to require $6*N^2$ searches. I.e. for each of the
$N$--points check all $6N$ tetrahedra whether they overlap that
location. This indeed would be very numerically inefficient.
Fortunately, this can be radically improved using a chaining mesh that
organizes tetrahedra contained in mesh cells or use tree structures
which group tetrahedra into regions of space they occupy. One could
use six trees, e.g., to organize bounding boxes containing tetrahedra
which is advantageous as they are generally smaller than the
circumsphere of tetrahedra which would require less storage. We,
however, implemented two other versions. One with a chaining mesh and
another that employs three bounding-box oct-trees offset from each
other containing disjoint sets of tetrahedra so that tetrahedra
cutting along tree node boundaries only do so typically for one of the
trees. We then parallelized the search with OpenMP and arrived at a
very practical tool to carry out the tessellation and measure the
quantities we were interested in.

For the most straightforward case where the stream density is given by
equation~(\ref{equ:rhostream}) the total density at a given locations is
now just the sum of these $\rho_s$ for all the intersecting
tetrahedra. This simple approach is what we used for the renderings in
the following section.

Since $\mathbf{y}$ will typically lie inside a tetrahedron, we can
also interpolate the primordial stream density to $\mathbf{y}$ if we
defined them by averaging over a cubical volume at the vertices of
this tetrahedron as described above.  A one-over-distance-squared
weighting (Shepard's method) works well.  Also in this case, the total
configuration space density is simply obtained by finding all
tetrahedra that contain the point under consideration and summing over
all their stream densities interpolated to this location. When
evaluating velocities, we interpolate the velocity field inside of a
tetrahedron from the velocities of the four particles that constitute
its vertices. Again, this is achieved by Shepard's method and no further
averaging is needed.

To be more explicit, let us emphasize here that all of what follows below is
obtained solely by post-processing existing simulations. No line
of code has to be changed in the readers' favourite cosmology code. As
long as the code writes out the particle IDs at every snapshot of the
simulation, and the connectivity of the initial particle distribution is known
or can be constructed, one can post-process simulations that already 
exist and measure, visualise and analyse it in new ways. 

With these definitions and implementation details in hand, we now
proceed to apply the method to a number of N--body simulations of
large scale structure formation. 


\section{Applications}

For our first applications, we chose to test the method on simulations
of the same physical volume, differing only in numerical resolution.

\subsection{$N$-body Simulations} We have carried out cosmological
$N$-body simulations of a volume of $40\,h^{-1}{\rm Mpc}$ length, run
with the tree-PM code {\sc Gadget-2} \citep{Springel:2005}.  The
initial conditions for these single-mass-resolution simulations were
generated with the {\sc Music} code \citep{Hahn:2011} keeping
large-scale phases identical with changing mass and spatial
resolution. We assume a concordance $\Lambda$CDM cosmological model
with density parameters $\Omega_{\rm m}=0.276$,
$\Omega_{\Lambda}=0.724$, power spectrum normalization
$\sigma_8=0.811$, Hubble constant $H_0=100\,h\,{\rm km}{\rm
  s}^{-1}{\rm Mpc}^{-1}$ with $h=0.703$ and a spectral index
$n_s=0.961$. The particle numbers, masses and force softenings of
these simulations are summarized in Table \ref{tab:sims}.  Very
clearly, this box is much too small for a careful statistical analysis of
the cosmic web. However, we chose it here to give us the opportunity
to study how our method converges at different mass resolution both in
the collapsed objects as well as in the cosmic web. A number of the
quantities we measure should hence not be understood as final
numbers/answers. Similarly, the highest resolution simulation we
discuss here takes very little computational resources. However, as we
shall see, these simulations will suffice to demonstrate the
advantages of the new approach.

\begin{table}
\begin{center}
\begin{tabular}{|c|c|c|}
\hline
number particles & $m_p / h^{-1}{\rm M}_\odot$ & $\epsilon / h^{-1}{\rm kpc}$ \\
\hline
\hline
$32^3$ & $1.50\times10^{11}$ & 100 \\
$64^3$ & $1.87\times10^{10}$ & 50 \\
$128^3$ & $2.34\times10^{9}$  & 25 \\
 $256^3$ &  $2.92\times10^{8}$ & 10 \\
\hline
\end{tabular}
\end{center}
\caption{The specifics of the suite of $N$-body simulations used in this
paper. All simulations are of a $40\,h^{-1}{\rm Mpc}$ cosmological volume,
$m_p$ is the particle mass, $\epsilon$ the force softening.}\label{tab:sims}
\end{table}

\subsection{Large Scale Structure \& Streams}

Our method allows to separate physically distinct structures.  The
number of streams at a given location can only ever be an odd number
as any fold will add two more streams to an existing one
\citep[e.g.][]{Arnold:1982, Shandarin:1989}. The notable exception is
at the location of caustics where points may sit such as to only
measure an odd number of additional streams. For the number of streams
defined at the particle positions, we can use this fact to select the
structures that are constituting the first caustic.
\begin{figure*}
\centerline{\includegraphics[width=0.9\textwidth]{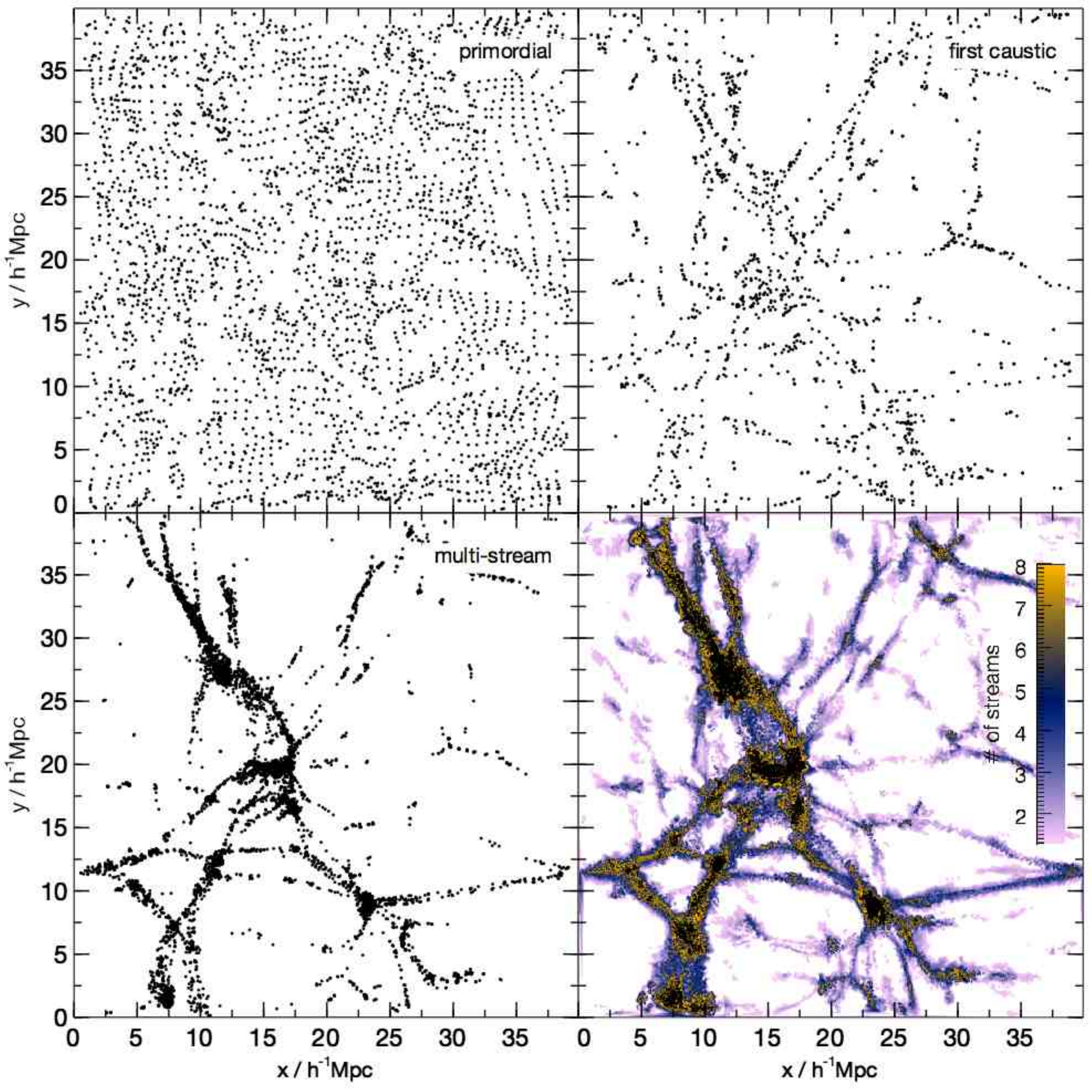}
}
\caption{Particles with different numbers of streams in a slice
  $0.5h^{-1}{\rm Mpc}$ thick. Top left: particles whose
  primordial stream does not overlap with other parts of the
  sheet. Top right: particles which are on their first caustic, 
  i.e. they measure a number of streams of two at their
  location. Bottom left: particles for which the number of streams is
  greater to or equal to three. Bottom right gives the average number of
  streams on an infinitesimally thin slice. Distinct physical
  components become clearly visible and separated. 
 }\label{fig:lss}
\end{figure*}
We illustrate the meaning of the local number of streams in
Figure~\ref{fig:lss}. The particles that record that they are part of
only their primordial stream clearly define the voids at the mass
scale that is resolved in the calculations. Particles that count two
streams surround the sheets formed between voids. When they undergo
their first caustic, they have already crossed through the sheet and
are turning around on the side opposite to from where they fell in
from.  The particles which measure three or more streams are also
shown and they trace the location of the collapsed objects well.  We
still consider all particles which count two streams or more as part
of collapsed objects. This is the same decomposition that can be made
in the GDE approach of \cite{Vogelsberger:2008} as shown for the
environment of Aquarius haloes in \cite{2011MNRAS.416.1377V} (their
Figure~B1) and \cite{Vogelsberger:2011} (their Figure~4). 

\begin{figure}
\begin{center}
\includegraphics[width=0.47\textwidth]{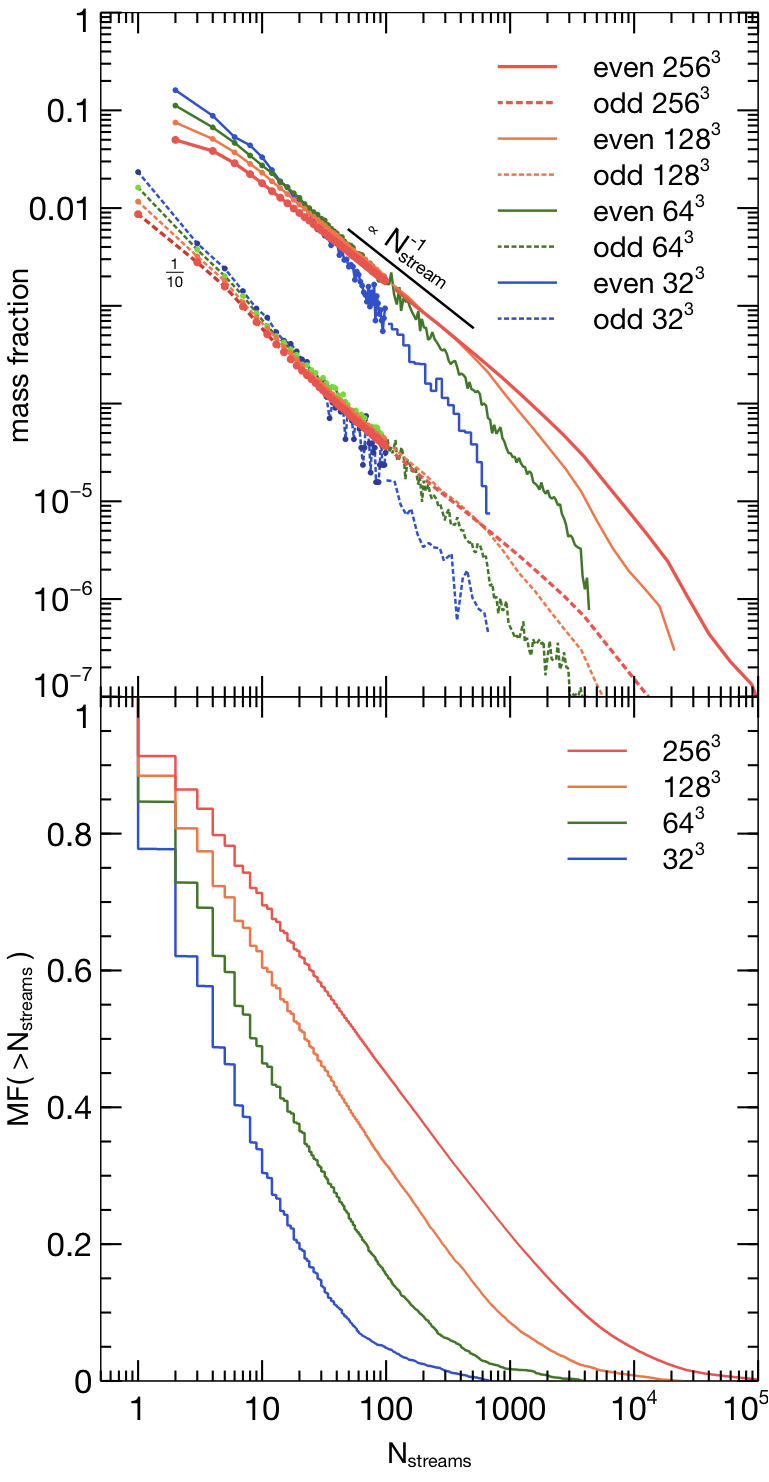}
\end{center}
\caption{The mass fraction distribution in streams. Many more odd
  numbers of streams are found than even ones. We show them as
  separate lines with the odd ones displaced down by a factor of ten
  for clarity. They differ by a factor approximately 4 and that
  ratio does not depend much on the resolution of the underlying dark
  matter simulation. An asymptotic slope of $\sim N_{stream}^{-1}$ develops for
  the higher resolutions for an intermediate number of streams. At low
  resolution steeper relations are inferred. For our most resolved
  simulations about 90 per cent of the mass is in collapsed structures. Some
  50 per cent of the mass is in locations with 20 streams or  more. }\label{fig:streams}
\end{figure}

We can see the distributions of the mass fractions as a function of
the number of streams in Figure~\ref{fig:streams}. We plot them for
particles recording odd and even counts separately. One may have
expected that the fraction of particles that are on caustics vs
particles that have odd numbers of streams to decrease, as the caustics are
better resolved for the high resolution runs. Instead, the offset
between odd and even numbered mass fractions is approximately
constant. This is just a feature of cold dark matter simulations that
with increasing resolution also more smaller objects can be resolved.
This is illustrated directly in the bottom panel of that Figure. The
cumulative distribution of mass above a given number of streams
clearly does not converge. This just reflects that there are many
small scale density fluctuations that collapse even earlier when the
simulations can resolve them. It is quite plausible that the total
fraction of collapsed mass will ultimately approach the very high
value of 99 per cent that has been estimated analytically by
\cite{Shen:2006} from the ellipsoidal collapse model.

\begin{figure}
\begin{center}
\includegraphics[width=0.38\textwidth]{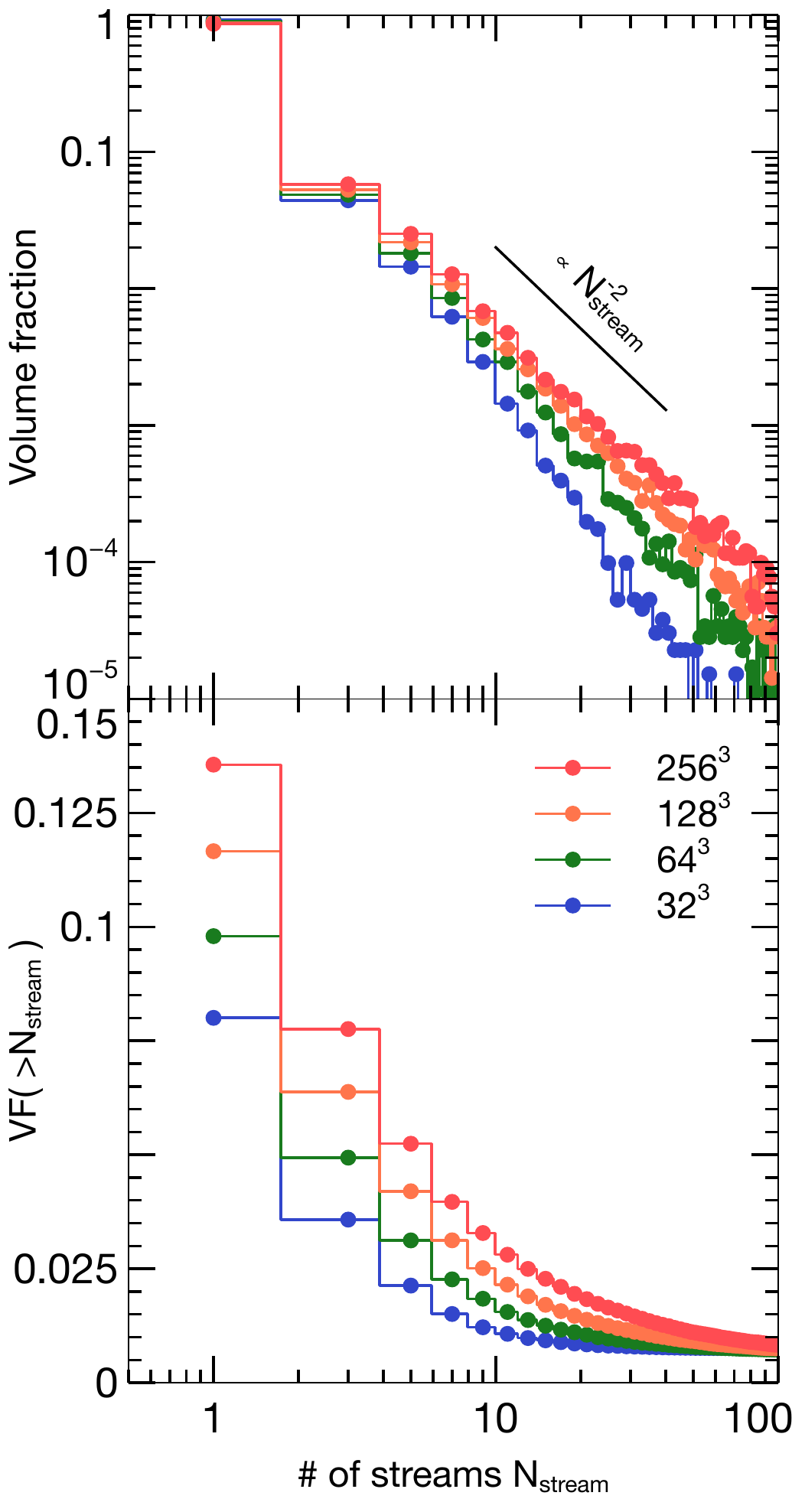}
\end{center}
\caption{The volume fraction distribution in streams; a resolution
  study. For our most resolved simulations about 85\% of the volume is
  in voids, around 7 (14) per cent is in collapsed structures
  ($N_{streams}$ larger or equal to three) for the $128^3$ ($256^3$)
  run. As expected in CDM, the fraction of volume occupied by
  collapsed objects does not converge. }\label{fig:streams_volfrac}
\end{figure}

For the volume averaged fraction as a function of streams and the
corresponding cumulative distribution in Figure~\ref{fig:streams_volfrac},
we observe a similar lack of convergence. Smaller and smaller pancakes are 
resolved as the resolution increases, making more and more volume
have had shell crossing in the past. 

It is clear from these results that questions about the shell-crossed
mass and volume fractions in cold dark matter simulations are
intimately tied to a scale. Only when introducing such a scale,
e.g. through filtering of density perturbations or a constant force
softening across resolutions, we could hope to obtain a meaningful
measure of these quantities.

This is compatible with previous results on mass and volume fractions
in the various parts of the cosmic web e.g. by
\cite{HahnPorciani:2007} using a fixed scale, or by
\cite{Aragon-Calvo:2010a} using adaptive filtering.  In both cases the
filtering scales are related to the non-linear scales today which is
the relevant scale for much of galaxy formation.

\begin{figure*}
\centerline{\includegraphics[width=0.97\textwidth]{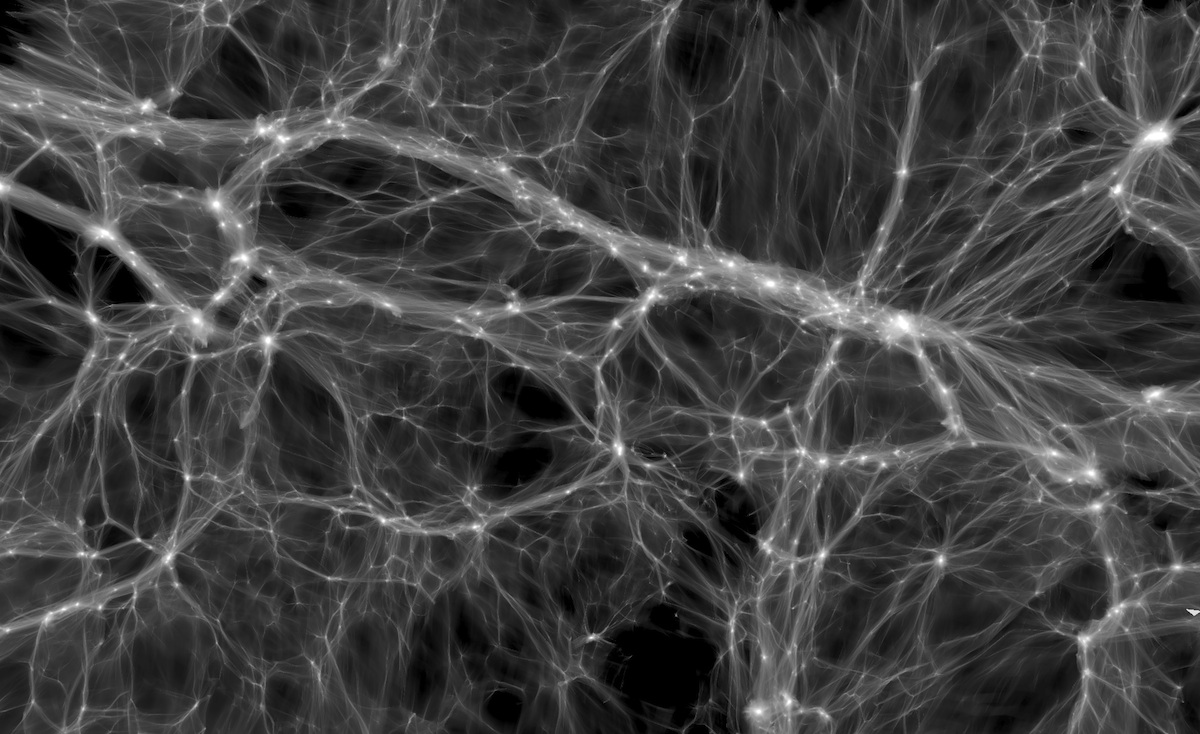}}
\caption{A rendering of the projected dark matter density in the
  $256^3$ run using our density estimator and our custom GPU based
  renderer.
  }\label{fig:pretty}
\end{figure*}

\begin{figure*}
\centerline{\includegraphics[width=0.47\textwidth]{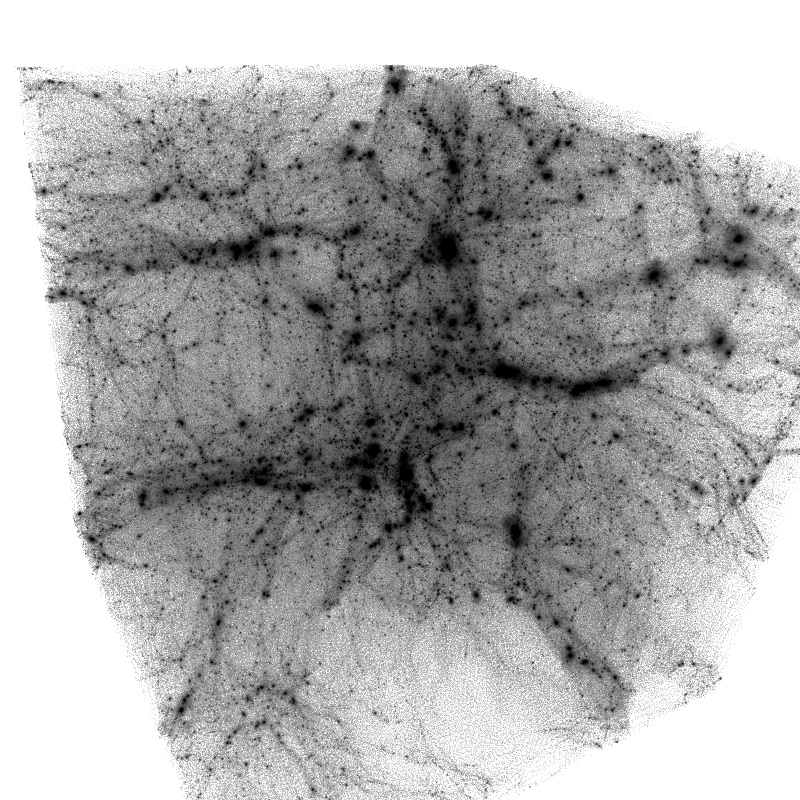} 
\includegraphics[width=0.47\textwidth]{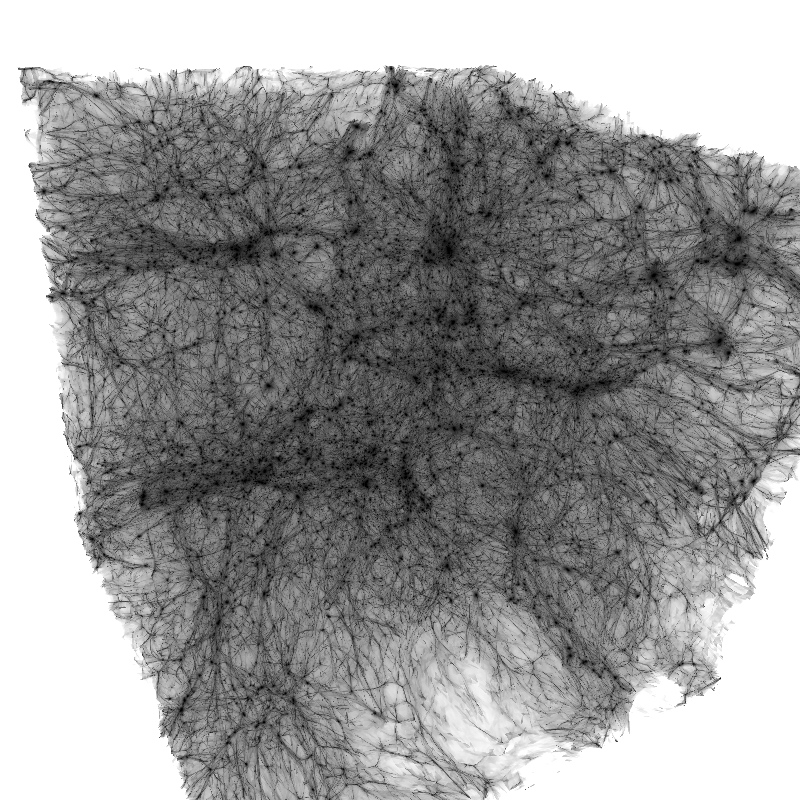}}
\caption{Comparison of the visual appearance of renderings of the dark
  matter density in the $256^3$ run using our new density estimator with a 
  simpler density estimate based on the log of the number of dark matter particles
  falling within given image pixels. While many of the well sampled
  regions are clearly apparent in both, the detailed structure of
  filaments, sheets and how they connect to voids becomes only
  apparent in our new approach shown on the right.
}\label{fig:pretty-box}
\end{figure*}

\subsection{Visualization}

The method presented here is also an ideal tool to visualize the
data from current N--body simulations and thus to further help extracting
physical insights from the calculations. Figure~\ref{fig:pretty} gives
an example that we have obtained with a custom-written OpenGL based
renderer of the tetrahedral mesh. The primordial stream densities
are averaged over abutting tetrahedra as described in the
implementation section above. Then tetrahedra are projected, taking
advantage of the OpenGL primitives designed specifically for polygonal
meshes. Details of this algorithm and variations as well as efficient
implementations of current graphics hardware will be
given in a forthcoming publication (Kaehler, Hahn and Abel, in prep.).

We compare this to the visual impression one obtains by plotting
individual points of a calculation vs. our new density definition in
Figure~\ref{fig:pretty-box}. One  can clearly see how filaments and
sheets in and surrounding voids can be distinguished easily now. The
visual impression is commensurate with the statistics we present next.

\begin{figure}
\centerline{\includegraphics[width=0.47\textwidth]{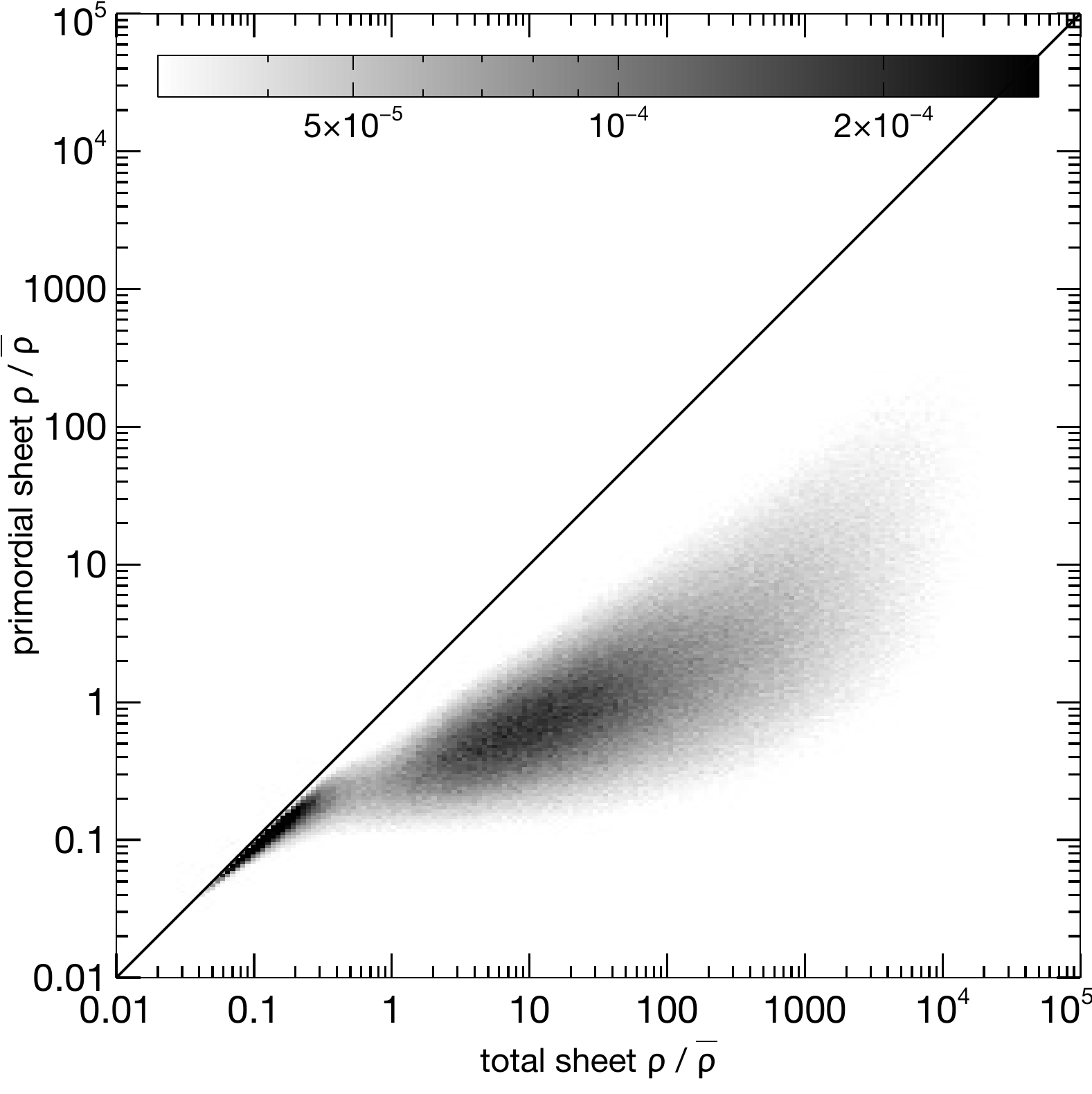}}
\caption{Two dimensional mass-weighted histogram comparing the density of the
  primordial dark matter sheet to the total sheet density in the
  $128^3$ simulation. Much of the mass is contained at configuration
  space densities about ten times the mean. The primordial stream
  density also scatters very much but its median density is close to
  the average density of the Universe. }\label{fig:primVSrhotot}
\end{figure}

\begin{figure}
\centerline{\includegraphics[width=0.37\textwidth]{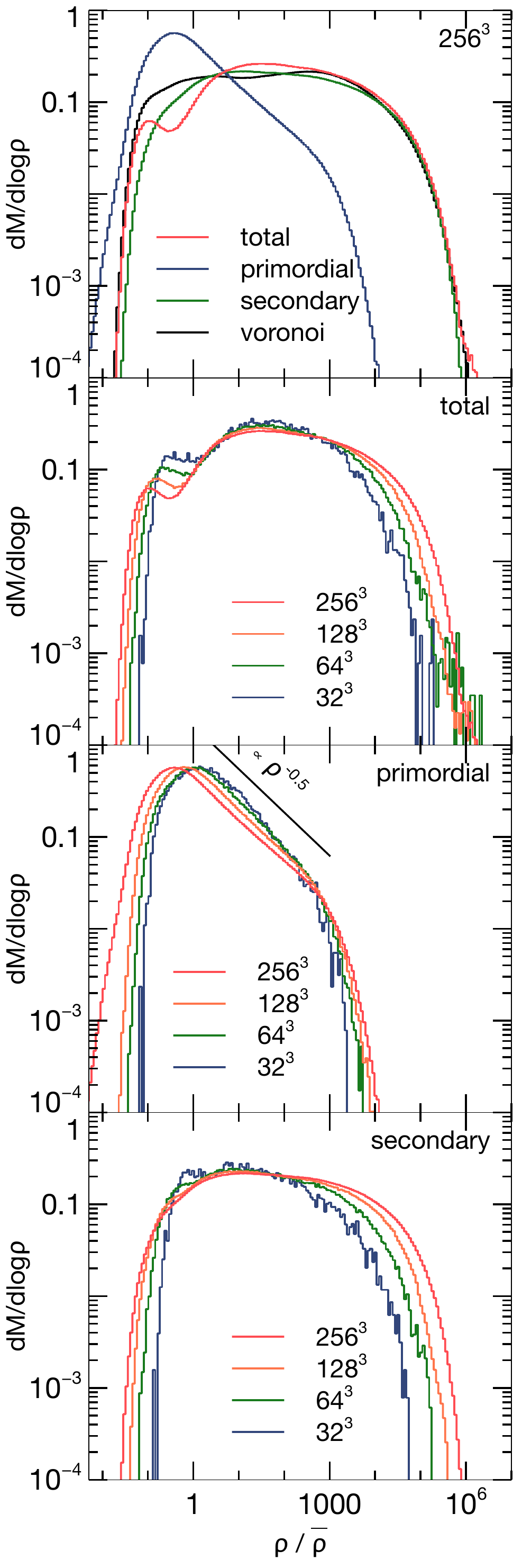}}
\caption{Mass-weighted density distributions.  The top left panel
  shows the histogram for four different densities defined at the
  particle locations for the $256^3$ run. The density estimated from a
  zeroth order estimate of the Voronoi tessellation, the total sheet
  dark matter density, the primordial stream density and the secondary
  stream density. The results of the resolution study is given in the
  other three panels. The total space density is given in the top
  right panel. The bottom left is the mass-weighted density pdf of
  sheet density in the primordial stream. The bottom right panel gives
  the density contributed by material that is not in the primordial
  stream.  }\label{fig:hist-rho-massw}
\end{figure}

\begin{figure}
\includegraphics[width=0.4\textwidth]{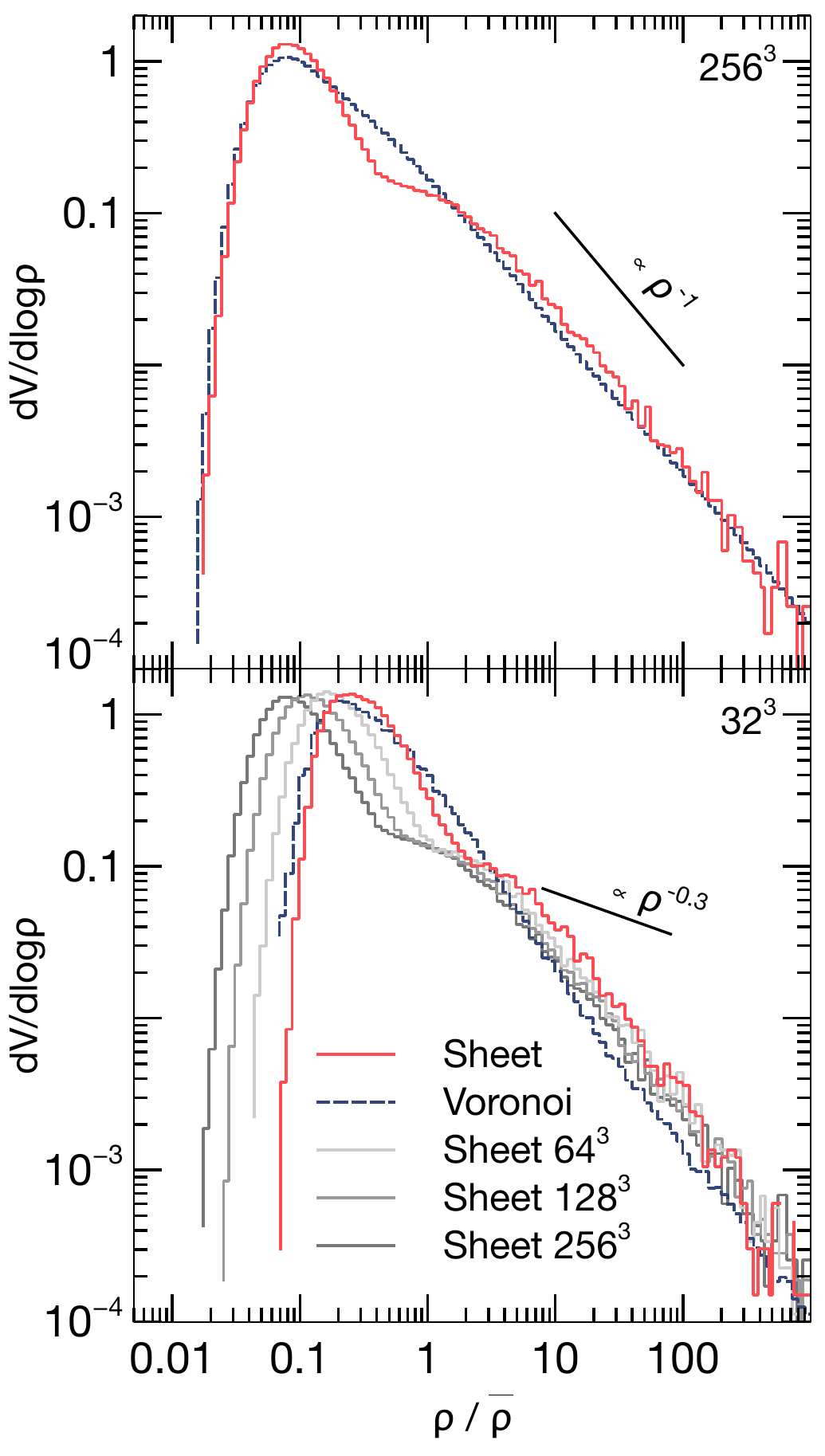}
\caption{Volume-weighted density distributions. The top panel shows
  the histograms for the $256^3$ run, the lower those for the $32^3$
  run. The zeroth-order density estimated from a Voronoi tessellation is
  shown with a dashed line, the total sheet dark matter density with a
  solid line.  At both resolutions, both the Voronoi and the stream
  density approach a $\rho^{-1}$ power-law at high densities. Also, the
  two methods produce different estimates at intermediate densities of
  $\rho/\bar{\rho}\sim10$. The bottom panel also shows in grey the
  density histograms from our method for all simulations to aid the
  comparison.}\label{fig:hist-rho-volw}
 \end{figure}
 
\subsection{Dark Matter Densities}

Figure~\ref{fig:primVSrhotot} gives the relation of the density in the
primordial stream to the total density evaluated at the locations of
all the particles, i.e. a mass-weighted histogram. At low densities
these are identical as this material is traced by the original sheet
and no folding has occurred. There is an enormous scatter at higher
densities which we can quantify
further. Figure~\ref{fig:hist-rho-massw} gives the mass-weighted
density distribution for all the simulations we have analyzed.  The
top panel summarizes the individual contributions for the $256^3$
simulation. The primordial stream density distribution peaks slightly
below mean density while the total mass weighted density distribution
is at much higher densities. The reason is that all the streams not
inside the primordial stream contributing to the density at the
location of the particle contribute much more to the total density at
high densities. That distribution is given by the green line in the
top panel labelled as ``secondary''. There is an apparent power law
part in the primordial stream densities visible. We will discuss this
further when considering volume-weighted distributions.  These are
given in Figure~\ref{fig:hist-rho-volw} where we show the volume-weighted dark
matter density. The total densities we estimate with our method are
labelled as ``Sheet'' . We also indicate the resolution of the dark matter
simulation used to compute it. The median of the stream density is
$1.2$ but its average is 26 times the mean. We also do not expect
these numbers to converge  as one continues to increase the
resolution.

\begin{figure}
\centerline{\includegraphics[width=0.47\textwidth]{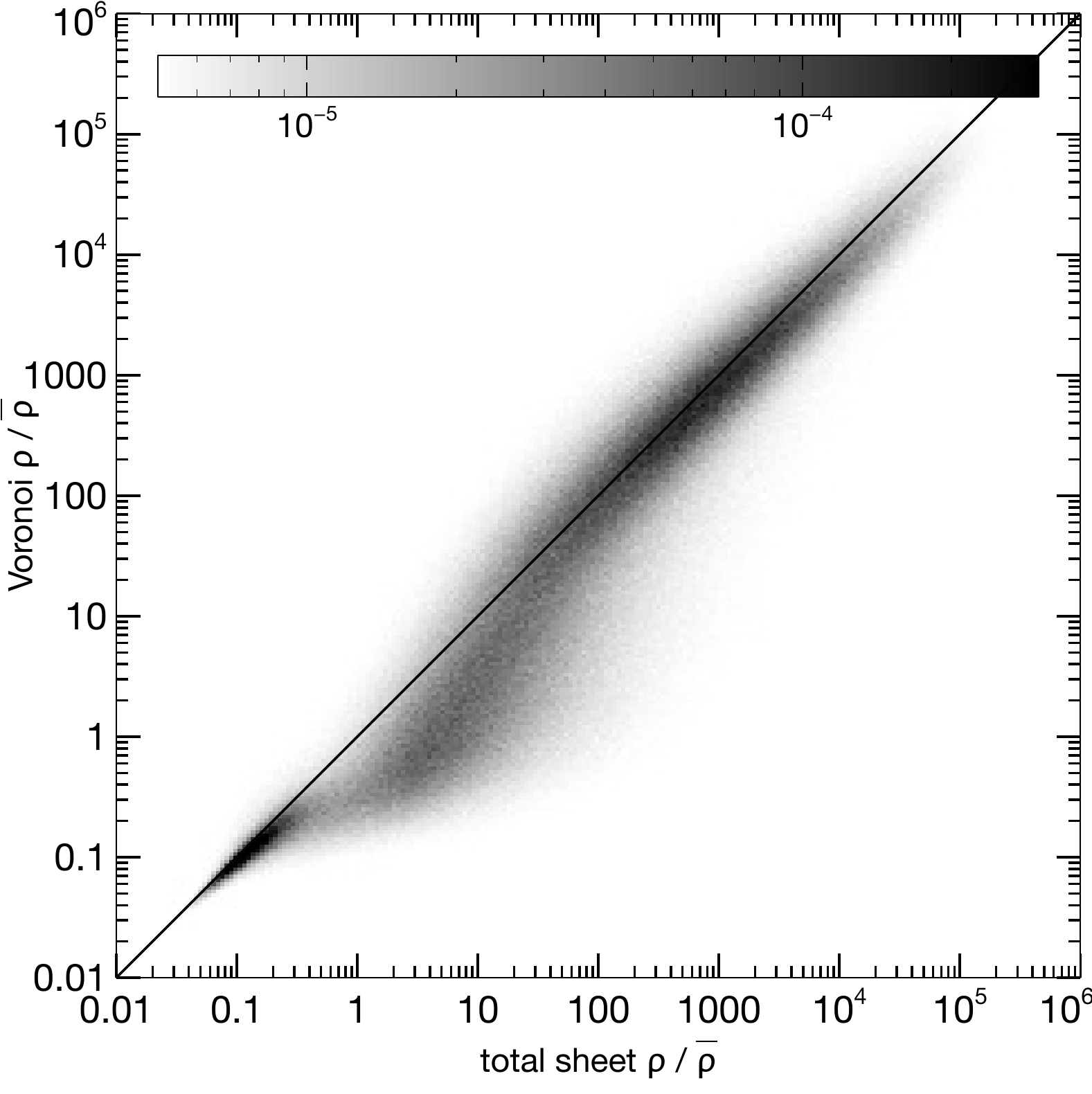}}
\caption{Two dimensional histogram comparing the zeroth--order Voronoi
  density estimate vs. the total sheet density. The correspondence is
  quite good. The largest difference is observed for values between a
  third and thirty times the mean density of dark matter. The
  zeroth--order Voronoi density estimators overestimates the volumes
  in regions around filaments and sheets. }\label{fig:vorVSrhotot}
\end{figure}

We also compared our new density estimates with the corresponding
results from another density estimator, which finds the unique Voronoi
cells around each particle. The density in that volume is then simply
defined as the mass of the enclosed particle divided by that volume
element. Following \cite{van-de-Weygaert:2009}, we refer to this as the
zeroth--order Voronoi density. Albeit, this is clearly more noisy than
the DTFE density estimator developed by \cite{Schaap:2000} and
\cite{Pelupessy:2003} since the density is defined for the smallest
region. DTFE is more advanced and employs averaging over nearby
tetrahedral cells.  These authors give comparisons of that estimator
to the smooth kernel estimates obtained with the otherwise popular 
Smoothed-Particle Hydrodynamics estimator. Like DTFE, the simple
zeroth--order Voronoi estimator we use tessellates the entire volume
and has no parameters and as such is a well suited benchmark for
comparison.

Quite strikingly, at low and high densities our density estimate is
very similar to the zeroth--order Voronoi volume based density
estimator (Figure~\ref{fig:vorVSrhotot}). Both methods do not
converge at the very low density tail when varying resolution. 
This is physical in that the simulations model smaller voids when 
the resolution is increased. These can achieve lower densities than 
their larger counterparts in lower resolution simulations. So, at higher
resolutions, it is not surprising to see a tail growing at those lowest
densities. Also, the peak distribution shifts continuously to lower
densities as more particles are employed.  Significant differences in
the volume fraction at a given density of our method with the
zeroth--order Voronoi method are seen at intermediate densities around
the mean density. This is understandable as the volumes that the
Voronoi tessellation provides, tend to connect particles in the voids
to the particles in the sheets and filaments. At that point it spreads
particles in filaments into volumes that are larger and consequently
estimates lower densities. In fact, these estimate are very
significantly different. So much so that, if we integrate the density
from our method over the volumes computed from the zeroth--order
Voronoi estimator, the total mass in the box is overestimated by more
than a factor of ten. It certainly would be interesting to compare our
density estimate not only to this zeroth-order Voronoi density but also
to other density estimators such as DTFE, adaptive Kernel softening,
etc.. This is, however, beyond the scope of this first exposition of
our approach.

Next, we will now apply our method to measuring a number of well studied
quantities in dark mater haloes.

\subsection{Radial profiles of haloes}

\begin{figure}
\begin{center}
\includegraphics[width=0.45\textwidth]{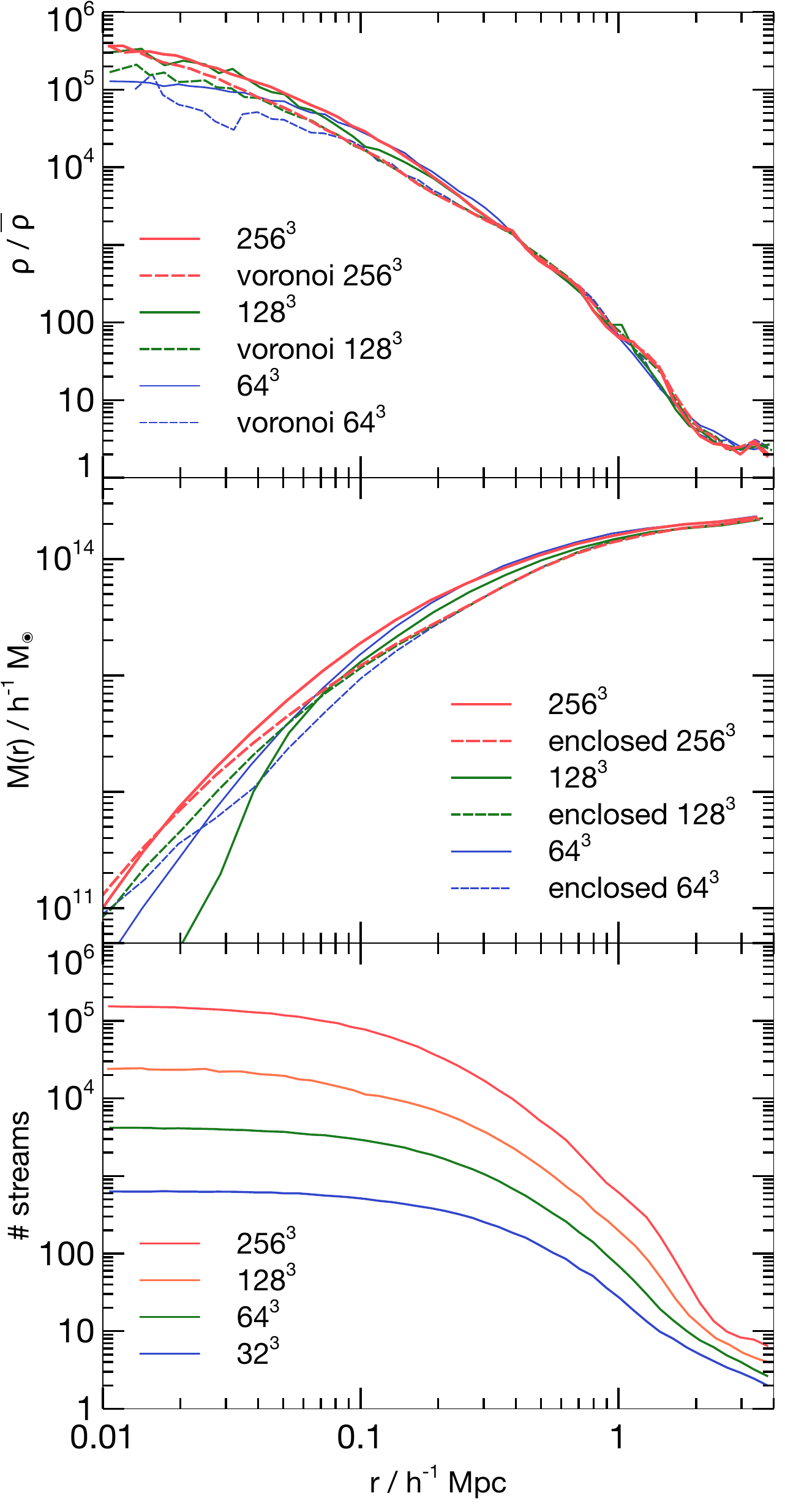} 
\end{center}
\caption{Dark Matter density profile in the most massive halo of
  $2\times 10^{14}M_\odot$ with $R_{vir}\approx 1{\rm Mpc}$ at
  redshift zero. The overdensity in the top, the enclosed dark matter
  (middle) as calculated from the density in the top panel, and the
  number of streams (bottom panel) are shown for all the different
  resolution simulations studied in this paper. The density profile
  estimated from our method starts to differ from the conventional
  estimate at scales as large as one third of the virial radius. As
  expected in CDM, the number of streams does not converge also in
  radial profiles as the resolution is increased.  }\label{fig:NFW}
\end{figure}

\cite{Navarro:1996} have discovered a universal radial profile of the
dark matter density in virialized haloes. This is one of the key
findings of cosmological N--body simulations and a large body of
literature has largely confirmed the finding. We will give these
profiles next. To get the best possible estimate, we chose 100,000
test positions and bin these in 50 radii,  spaced
logarithmically in radius.  Figure~\ref{fig:NFW} summarizes our
findings.  The density profiles computed from the dark matter sheet are
somewhat shallower and have about 50\% larger central densities at all
resolutions for the single halo we have analyzed. Physically, it is
conceivable that volume elements formed by particles on radial orbits
oscillate such that the bounding regions have a higher probability to
be found at large radii while still contributing to the density
interior to small radii.  Similarly, one can picture particles of the
sheet orbiting the center at larger radii such that the volume element
they span can contribute to the center. Our method has a well defined
density at all radii and it is bound to be a constant at the lowest of
radii where one averages always over the same tetrahedra. At large
radii, the new density estimate and the Voronoi estimates all agree
extraordinarily well.  This is true even in the infall region. The
Voronoi estimates at all radii are perfectly consistent with a simpler
estimate based on the particle mass binned in shells divided by the
shell volume (not shown).

The masses included within a radius do converge also quite well with
our estimate being consistently slightly higher at all radii. 
This is understandable as mass from particles outside a given
radius can contribute if they span a volume element that has nodes
inside the radius.

The number of streams that contribute to the profile are given in the
lower panel of Figure~\ref{fig:NFW}. Not surprisingly, these increase
with increasing resolution. If one scales them by factors of eight
between increasing resolutions, some closer convergence is
observed. Between the $128^3$ and $256^3$ simulations, there remains a
difference of about 30\% which is likely just due to streams
contributing from larger radii to the regions inside the particles
spanning the volume element.

It certainly would be interesting that the dark matter density profile
in the central parts of haloes could be different than one estimates
by measuring dark matter particles inside a given radius.  It is
also suggestive how well our density profiles converge from $128^3$ to
$256^3$ particles.
However, if the mass profile were indeed different, also the forces
contributing to the particle motions would be changed. So even if our
density estimator were more accurate, one could not prove that the
result shown in Figure~\ref{fig:NFW} is the correct physical one until
one has evolved the dark matter sheet consistently, i.e. using
accelerations created by the density distribution of the actual sheet
elements. We discuss some potential approaches to carry out such
simulations in the discussion section.

\subsection{Velocity dispersions and the dark matter ``entropy''}

\begin{figure}
\begin{center}
\includegraphics[width=0.45\textwidth]{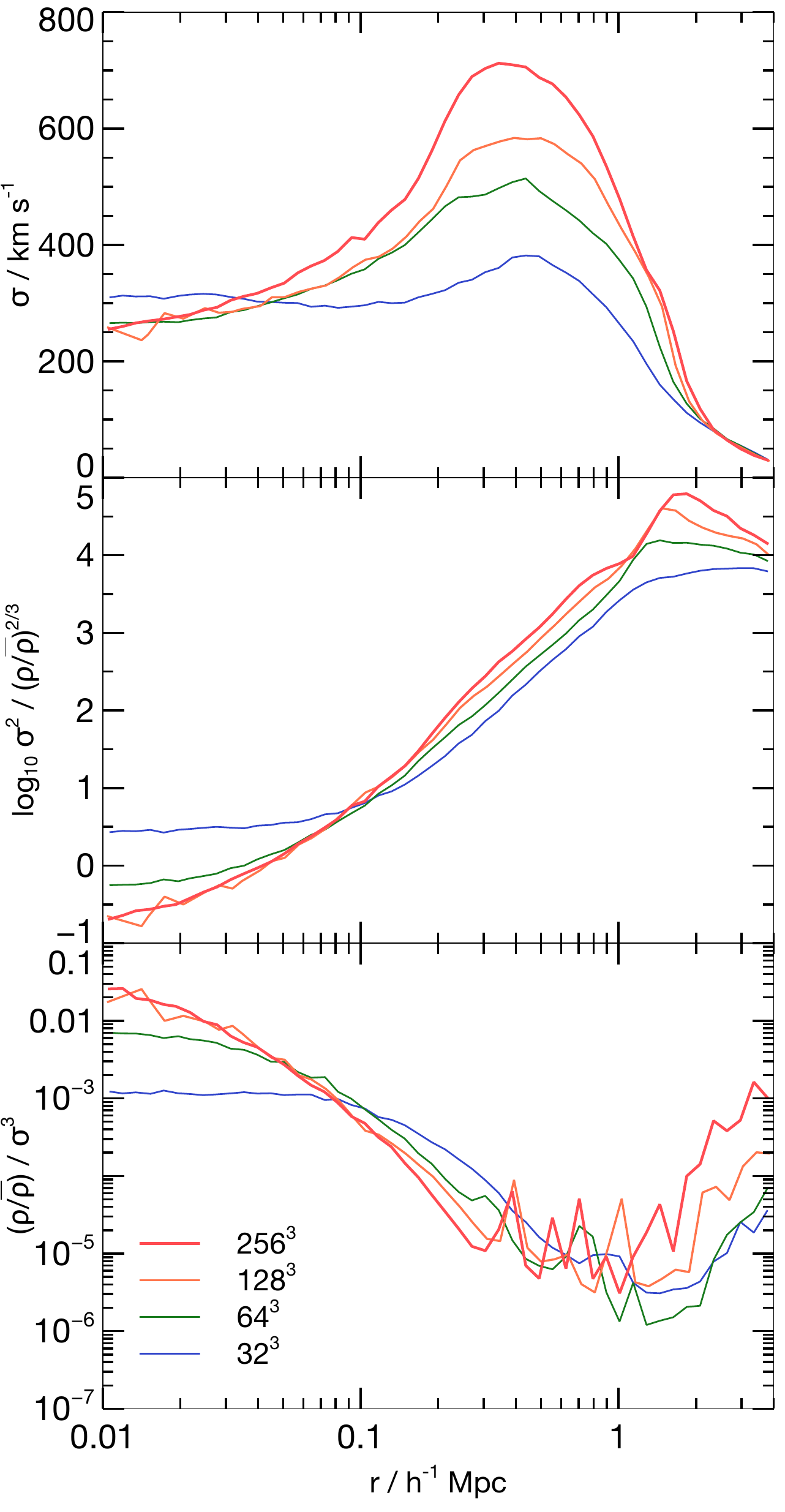} 
\end{center}
\caption{Radial spherically averaged profiles of the velocity
  dispersion (top), the dark matter ``entropy''
  $\sigma^2/(\rho/\bar{\rho})^{2/3}$ (middle) and the pseudo
  phase-space density (bottom) for the same halo as in
  Figure~\ref{fig:NFW}. The velocity dispersion is remarkably flat
  inside about one tenth of the virial radius. The dark matter
  ``entropy'' profile also shows signs of already converging at the
  modest resolutions employed here. Using the microscopic velocity
  dispersion of our approach which removes the bulk flows does not
  give the typical powerlaw behavior in the pseudo-phase-space
  density found when using the total dispersion of particles at those
  radii. 
 }\label{fig:veldisp}
\end{figure}

While our method gives access to the full fine-grained phase-space
structure, we chose to only show moments to serve as an example of what
the method is good for, and to be able to compare to work done on
this previously. While the collisionless fluid does not experience
microphysical collisions, the scattering provided by the time-varying 
gravitational potential leads to mixing in phase-space. 
The velocity dispersion of the particles is a measure of the effective
pressure of the dark matter, which is of relevance for understanding
the dynamical structure of orbits, i.e. e.g. the expectations of how observable
stars move in the DM potential. 
Figure~\ref{fig:veldisp} summarizes the radial profiles of the
velocity dispersion for the same halo we have analyzed above for 
density profiles. We again draw particles at random positions in
spherical shells for which we measure the stream-density-weighted
bulk velocity, subtract it from the stream-density-weighted local
velocity dispersion, before we finally average it to obtain the velocity
dispersion in radial shells. 

These velocity dispersions  differ quite significantly from the
similarly termed quantities presented previously \citep[e.g.][and
references therein]{Navarro:2010} . While this may seem surprising, one
has to keep in mind that we measure the dispersion at a single
point, i.e. we do not carry out any averaging over volume. 
Dispersions quoted in the past measured a combination of a bulk and
local dispersion. This will not only have large sampling error but
also confuse turbulent bulk motions with actual microphysical
velocity distributions. Indeed, we can see that our measured velocity
dispersion does not converge at scales of about one half the virial
radius, as  only about one thousand streams contribute there for our
highest resolution results. The distribution functions at this location will be
quite anisotropic and a single temperature will be a bad fit (see
below).  The halo is remarkably cold in the center -- having less than half the
velocity dispersion expected from the virial velocity. In the same
figure, we also show the pseudo phase-space density of
\cite{Taylor:2001} which has been found to be a perfect power-law
entirely independent of resolution \citep{Navarro:2010}. When we
measure the average of only the fine-grained quantities, as shown in
Figure~\ref{fig:veldisp}, this perfect power-law disappears. This may
suggest that much of the measure is dominated by large-scale bulk
flows. It is worthwhile to explore this further with higher resolution
simulations, where one can more confidently separate thermal motions
from the bulk velocity dispersion. 

\begin{figure}
\begin{center}
\includegraphics[width=0.47\textwidth]{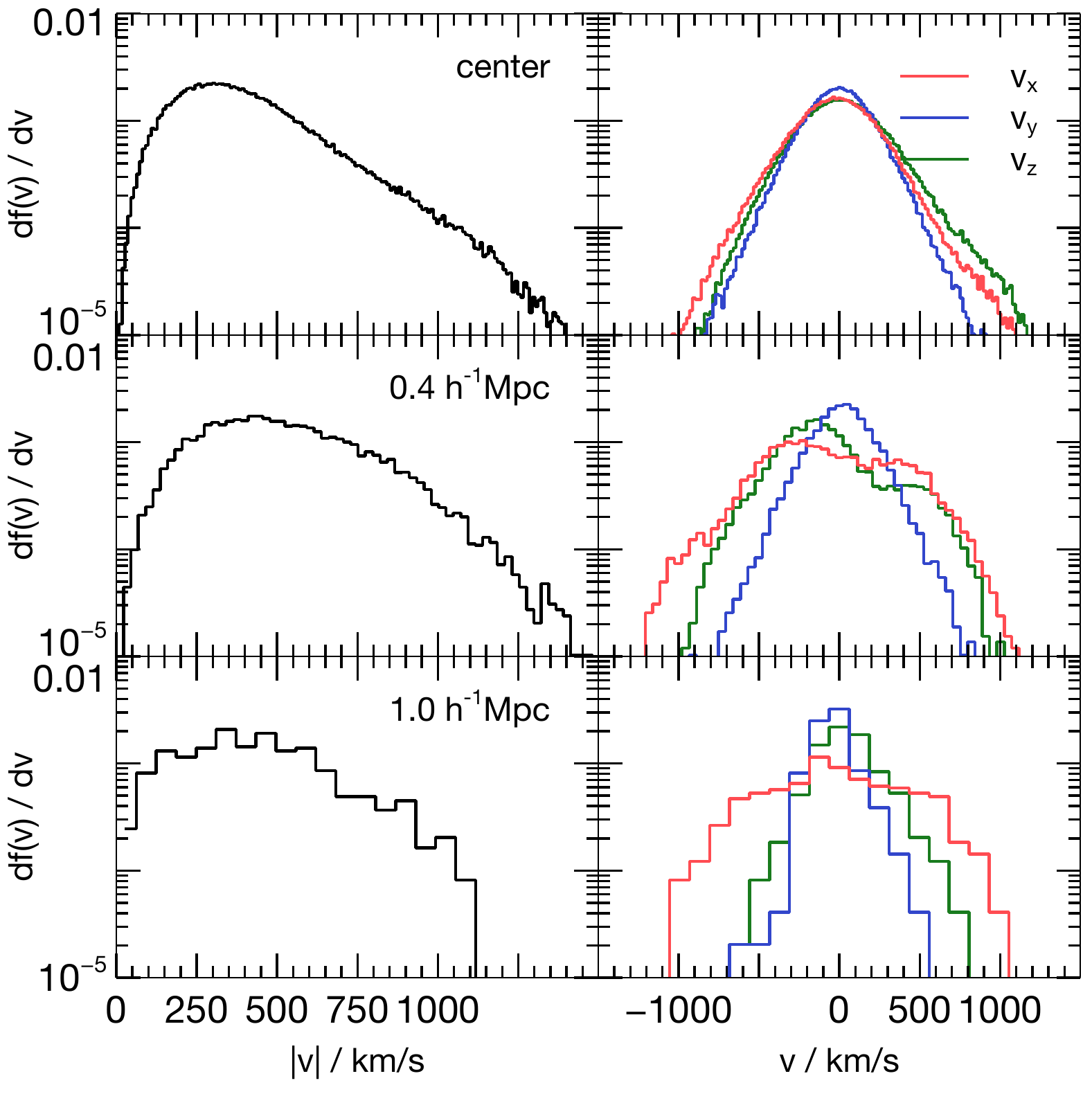} 
\end{center}
\caption{The velocity distribution function binned in velocity space
  at three points at varying distance from the center of the halo
  (left panels) and the individual distributions of the velocity
  components (right). We used 200, 80 and 30 bins for the points at
  the center, 0.4 $h^{-1}$Mpc and $1 h^{-1}$Mpc from the center along
  one axis. These had 150,000, 13,000 and 400 individual streams that
  contributed.  }\label{fig:violent}
\end{figure}

\begin{figure*}
\begin{center}
\includegraphics[width=0.97\textwidth]{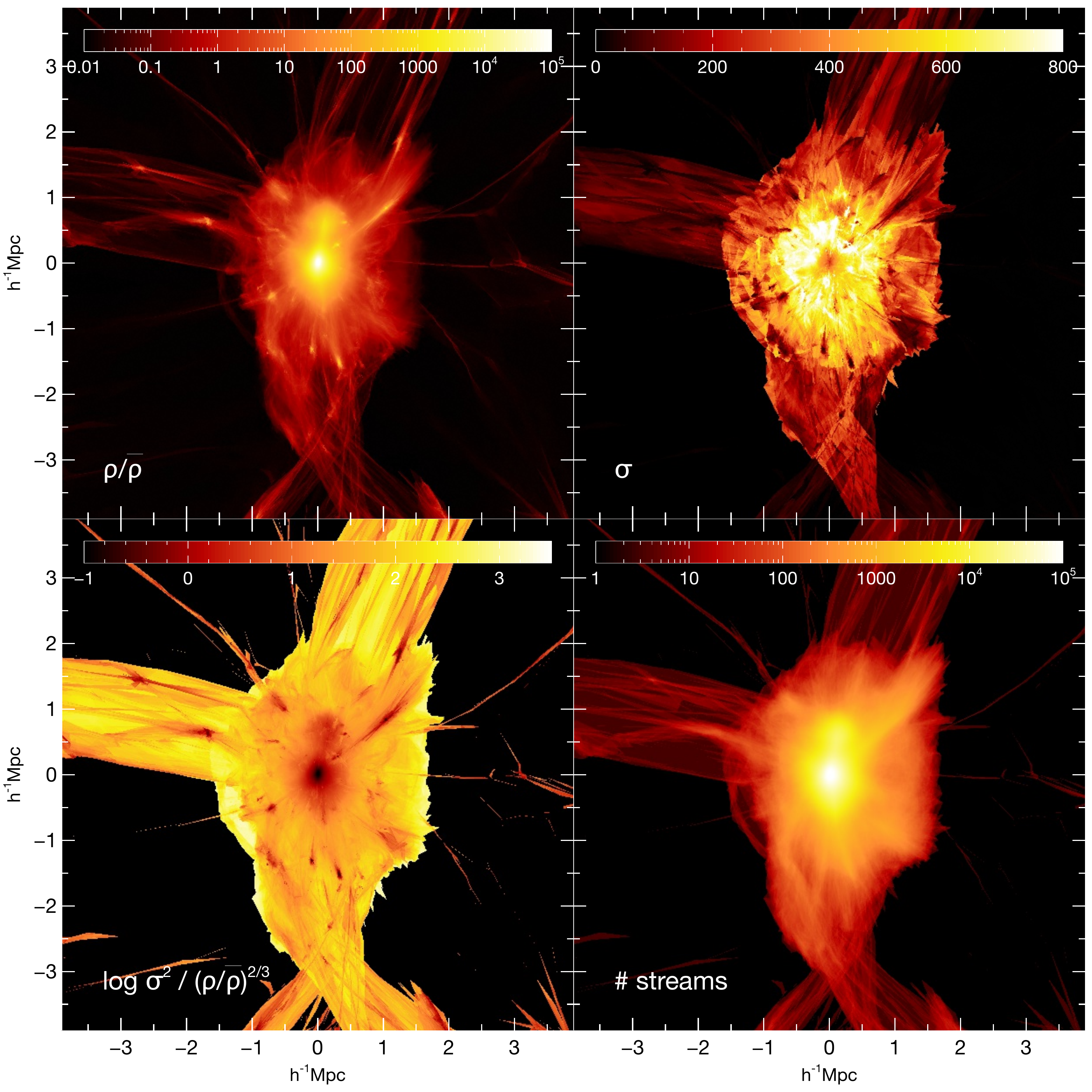} 
\end{center}
\caption{Infinitesimally thin slice through the $256^3$ simulations
  for the most massive halo.  We show the density in units of the mean
  density (top left), the stream-density weighted velocity dispersion
  in kilometers per second (top right), the dark matter entropy
  [(km/s)$^2$], computed from the density and stream weighted velocity
  dispersion $\sigma^2/(\rho/\bar{\rho})^{2/3}$, and the average number
  of streams (not stream density weighted) at the bottom
  right. Clearly, our approach gives information so far thought
  inaccessible from current simulations.  }\label{fig:clusterpanels}
\end{figure*}

It is though just as interesting to check the actual distribution
function of dark matter velocities at a given point. The seminal work
of \cite{Lynden-Bell:1967b} discussed this in the context of stellar
systems. The global distribution has been measured from simulations
many times \cite[see e.g.][]{Hoeft:2004, Wise:2007a, Navarro:2010,
  Vogelsberger:2009} but, to the best of our knowledge, this was never
done at individual points in the simulations. Figure~\ref{fig:violent}
summarizes the distributions found at three different points in the
most massive halo. We show it at the center where a relatively hot
component overlays a colder one. At the center, the distributions of
the individual velocity components have peaks that almost coincide and
widths which are quite similar as well. They are not too far from an
isotropic Maxwell-Boltzmann distribution in their cores. As we step
out in radius, the situation changes rapidly and the microphysical flow
structure clearly shows more and more anisotropy. Interestingly, a
quite hot component is seen along the $x$ direction. At the same the
velocity distribution in the $y$ direction is the coldest at all
radii. These distributions are consistent with the visual impression
obtained from the velocity dispersion slice in 
Figures~\ref{fig:clusterpanels} and \ref{fig:veldisp} which also shows
that some of the hottest DM fluid elements are found just inside the
virial radius.

It is remarkable how much physics can be learned from even these low
resolution simulations analyzed here. For the halo we just
discussed, there are not even 600,000 dark matter particles inside the
virial radius of $1.4h^{-1}{\rm Mpc}$. At the same time, there are
already enough streams to compute meaningful measures of the structure
of phase-space. We are certainly looking forward to carrying out a
more detailed analysis on higher resolution simulations. This point is
born out by the visual impression given by infinitesimal slices as shown in
\ref{fig:clusterpanels}, which we will describe next.
 
\subsection{Slices of Density and Dark Matter ``Entropy''}

To aid in the interpretation of the profiles we have just presented,
we also give two dimensional slices through the dark matter density
and ``entropy'', which we define analogously to the adiabats used when
studying e.g. galaxy clusters hydrodynamically simply as $S_{DM} =
\sigma^2/(\rho/\bar{\rho})^{2/3}$, which then has units of the
square of a velocity. Also, the average number of streams contributing
to every point on the slice is given. Material from the voids falls in
perfectly cold. We can think of the velocity dispersion as a measure
of the temperature of the fluid. It is ill defined in the single
stream regions falling from the voids. However, these carry very
little mass. Then we see a region that extends to about two Mpc from
the center which hosts multi-stream material of the order of about ten
streams. The virial radius, which is approximately at one Mpc, shows a
marked increase in the velocity dispersion and a much smoother density
structure. Even on this scale, we can see the cold central isothermal
part of the object, both in the velocity dispersion and in the
entropy. Substructure is easily seen as cold low entropy material
embedded in the hot halo. Many of the structures seen here are very
reminiscent of adiabatic hydrodynamical simulations of galaxy clusters
\citep[e.g.][]{Frenk:1999} and even first star formation \citep{Abel:2002c}
where gas enters haloes predominantly through filaments and shock
heats, resulting in a halo with rising entropy profiles with radius.

\subsection{General interpolation to any point in space}

There are large advantages to have well-defined grids which allows one
to interpolate to any point in space. This is a very obvious
observation of course, it is, however, a large step forward in
understanding dark matter simulations. This has led to a number of
approaches being devised that allow such interpolation, such as the
methods discussed in the introduction. Here we discuss but a few
approaches on how to use the tessellated dark matter sheet for
interpolation.

\begin{figure}
\begin{center}
\includegraphics[width=0.47\textwidth]{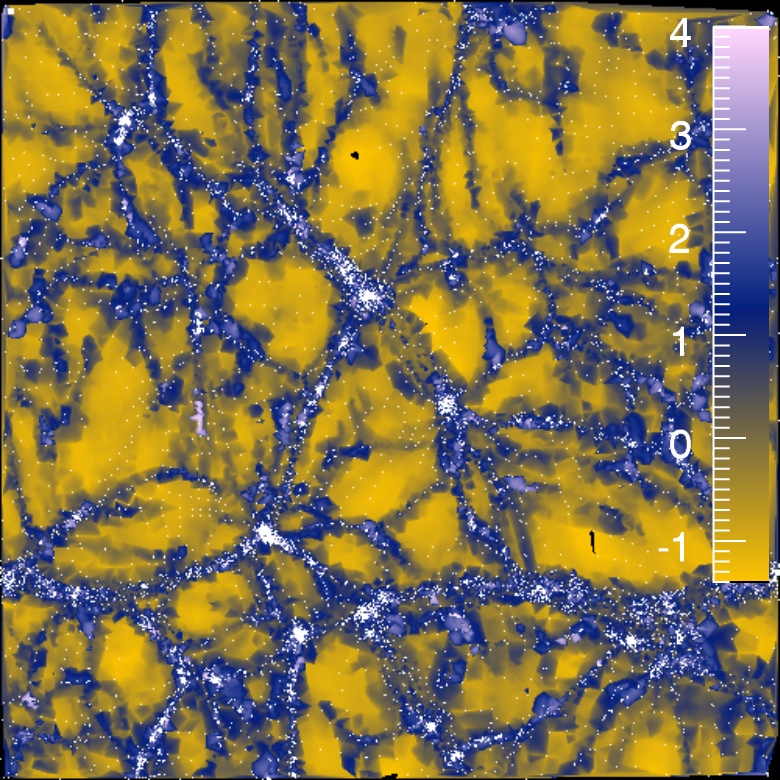} 
\end{center}
\caption{The logarithm of the density in an infinitesimally thin slice
  in units of the mean density for the $128^3$ simulation. The white
  dots show the location of the particles which would contribute to a
  cloud in cell interpolation on a grid with cells as large as the
  mean particle spacing. The squares at the top left show the area to
  which these individual particles would contribute to at their
  locations. 
}\label{fig:DensityVsCIC}
\end{figure}

When probing the sheet at the particle locations, we find the
primordial stream densities, total space densities, number of streams,
velocities etc.. Hence, we have sampled the full volume and have a
non-uniform distribution of the fields we aim to interpolate. One may
choose to achieve further interpolation by using a distance-weighted
estimate from the nearest particle locations. An efficient way to
find two dimensional slices is to take all tetrahedron edges and
compute their intersections with the plane to be interpolated
to. Along every edge one can now linearly interpolate the values of
the vertices to the plane. The resulting scattered data on the plane
is then triangulated again and interpolated with linear interpolation
between nodes. As an example the slice of the total sheet density is
shown in Figure~\ref{fig:DensityVsCIC} which also gives a visual clue
to how cloud in cell interpolation would sample the density field. 

\begin{figure}
\begin{center}
\includegraphics[width=0.47\textwidth]{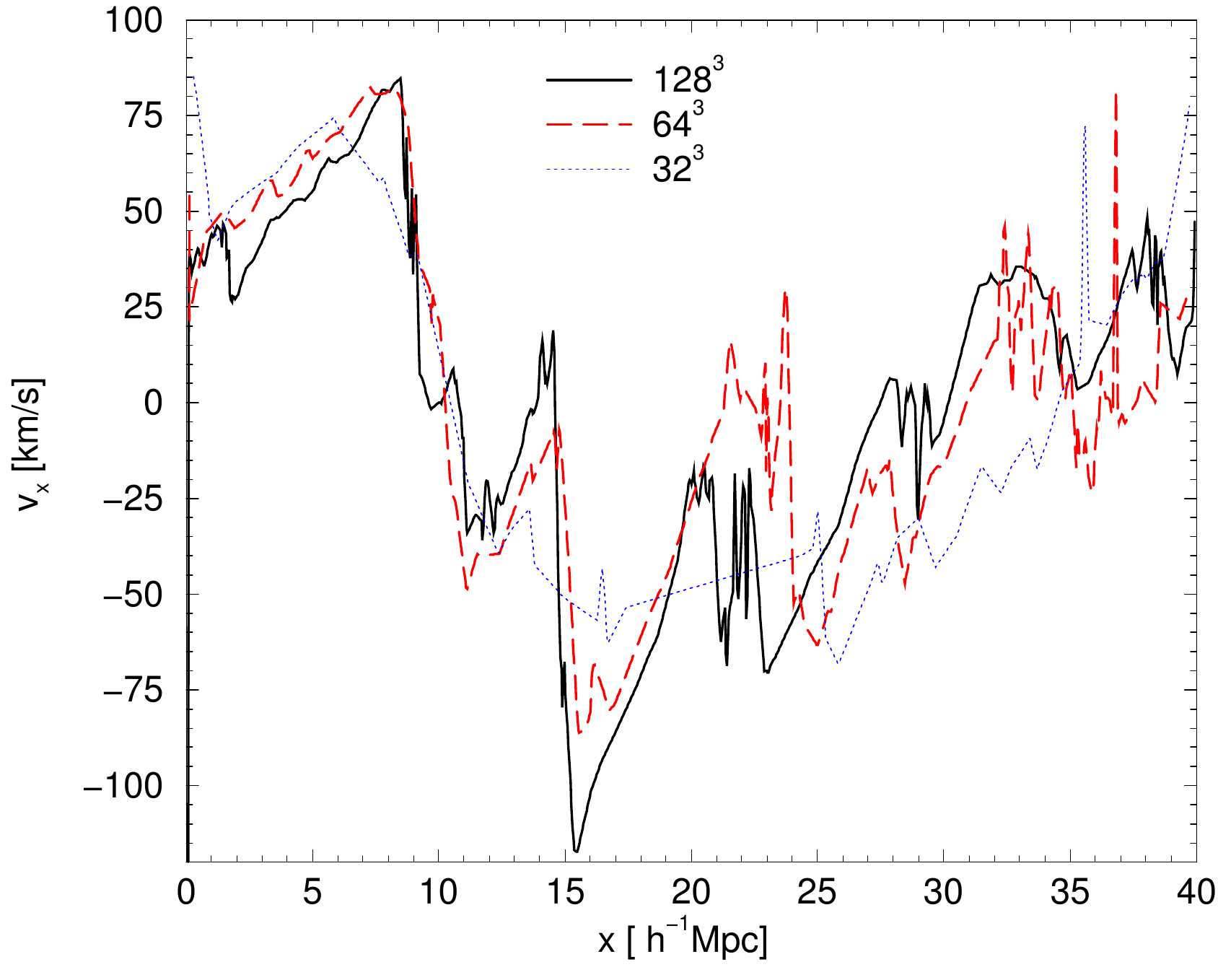} 
\end{center}
\caption{The velocity field along a one dimensional line extracted
  using tetrahedron edges to interpolate to a slice plane. The
  differences in resolution are understandably quite large given the
  large range of mass resolution of the simulations. The large scale
  features remain recognizable even at the lowest of resolutions.
The ``N-shaped'' infall regions are seen for many structures. 
 }\label{fig:velSkewer}
\end{figure}

Similarly, this  allows us to extract one dimensional skewers at
arbitrary resolution from  N--body simulations. As an example,
Figure~\ref{fig:velSkewer} compares the velocity along a random line
through the volume for different resolutions. The large scale modes are
all consistent by design. It is interesting to see that convergence is
quite slow and suggests to extend this analysis rigorously to much
higher resolutions.


\section{Discussion}

\cite{Vogelsberger:2008}, \cite{White:2009} and \cite{Vogelsberger:2011} developed the
GDE formalism to allow a calculation of the primordial stream
density. Their approach modifies the simulation code to integrate an
evolution equation of the tidal tensor along with every particle
trajectory.  In principle, this can be much more accurate since the
local stretching of the dark matter sheet is calculated at every time
step of the calculation. It will be of great interest to compare our
approach to theirs in detail. This will require to carry out the
calculations with both methods on an identical simulation to
facilitate a particle by particle comparison. This should be
particularly interesting given that our method can, in addition to the
primordial stream density, also provide the total space density and
number of streams at every location. Since our approach also gives
full fine-grained phase-space information, it seems plausible one
could combine both approaches to a hybrid which inherits the advantages of
both.  

More generally, both the GDE and our approach suggest a number of
possible approaches to improve the accuracy of N--body calculations.
Almost all current cosmological N--body solvers employ the particle
mesh method at least for the largest scales in the calculation. The
cloud in cell approximation is used to interpolate the dark matter
particles to a grid on which the gravitational potential will be
evaluated before differencing it to obtain the gravitational forces on
the particles. Since one integrates twice to get the potential from
the density field and only differentiates once, this method gives
reasonably smooth gravitational forces. However, it inherently models
a very noisy inaccurate density distribution obtained from CIC which
will have the largest relative errors in poorly sampled regions such
as voids, pancakes and filaments (see Figure~\ref{fig:DensityVsCIC}).

We have shown that our density estimator would have significantly more
fidelity and reliability for these large regions. It is in principle
quite straightforward to modify an existing particle mesh code to make
use of our density estimator and then derive more accurate potentials and
forces from it.  It only involves the interpolation step to the
grid. When interpolating the contribution back to the particle positions
one could make use of the known analytical solutions to the Newtonian
potential of homogeneous polyhedra \citep{Waldvogel:1979}. Similarly,
these analytic formulae could be applied in direct summation and
tree-based codes. A priori it may seem difficult to imagine how to
construct trees efficiently when considering that the tetrahedra may
become exceedingly distorted and elongated and would cover many nodes
of the tree. However, any new code would most likely ever only be
employed using local mesh refinement given that the tessellation we
suggest gives many opportunities to discover the regions of the flow
which may be prone to errors. The local curvature of the flow
compared to the tetrahedra edges is one measure but also the axis
ratio of individual tetrahedra provides an estimate where the flow
would benefit from refinement. The key to any such new method will
have to be to fully consider the dark matter as a fluid so that spurious
particle-particle interactions may be avoided and multi-mass
resolution becomes feasible. Given a locally refined mesh, tree
structures will remain useful in rapidly finding neighbours and retain
$n\,\log n$ scaling.

There are a number of improvements possible that will help to develop
our GPU assisted volume rendering further. Using vertex values and
some form of Shepard's method to carry out distance dependent
weighting should still likely be very fast on current GPUs even when
drawing billions of tetrahedra. 

Higher order interpolation in fact could be another avenue to improve
on the method suggested here. We have only implemented the very
simplest of ideas. Namely that volume elements in phase-space are
uniformly filled with the dark matter fluid. This is similar in spirit
to donor cell methods used decades ago for hydrodynamical flows. We
believe that it will be possible to improve on our approach
significantly.  Up-winded schemes with linear reconstruction were a
large gain in accuracy in numerical fluid dynamics and similar
improvements are certainly possible here.

Given that one now has a natural grid that can be used to interpolate
any state variables as well as the full fine-grained phase-space
structure, one can also define differentials on it. Consequently, it
becomes possible to study vorticity, divergences as well as carry out
the Cauchy-Stokes decomposition of the dark matter velocity
fields. This way one can separate bulk, shear and rotational 
 components of the velocity fields which undoubtedly will
make it possible to track down the physical origin of dark matter
density profiles as well as to better understand the internal structure
of haloes and the cosmic web. 

There is a remarkably large number of applications where we think our
method can aid to gain new insights. Whether it is gravitational
lensing to find more accurate lensing potentials to studying the origin
of the angular momentum profiles\citep[see e.g. Fig. (12)
in][]{Bullock:2001a}. Obviously the connection between dark matter and the
baryons they host can be explored much more fully now as well.

As we were preparing this manuscript, \cite{Shandarin:2011} posted a
paper on the electronic preprint server which explores the same basic
idea as the one we present here. The concept of tessellating
phase-space and tracking the dark matter sheet is identical to
ours. Details of the implementation and what to think of as
fundamental parts of the approach are not the same.  Their choice of
tessellation is quite different. 
They pick the minimal combination of tetrahedra of the unit cube
possible which has five elements where one of them is twice the volume
of the other four and itself does not tesselate the space
uniformly. Consequently they alternately rotate adjacent cubes such
that the edges of tetrahedra never cross. This is effective albeit
likely more cumbersome for a practical implementation. The powerlaw
\cite{Shandarin:2011} gives for the volume fraction as a function of
the number of streams ($f_V(N_{stream}) = 0.93\,N_{stream}^{-2.82\pm
  0.05}$) is to be compared with our
Figure~\ref{fig:streams_volfrac}. Their power-law fits approximately
our $32^3$ run and likely just reflects the fact that the single
simulation they study had approximately two times worse mass
resolution than our $32^3$ run with an effective gravitational
softening length about five times larger than ours. So both approaches
do agree. We at this time would not attach a special meaning to this
power-law as it clearly is strongly resolution dependent with our
highest resolution run giving something close to $N_{stream}^{-2}$.
Our description also discusses the fine-grained structure in the
velocity directions of phase space, discusses halo properties and
profiles and gives visualizations of the dark matter density not given
by \cite{Shandarin:2011}.

\section{Summary}
We presented a novel approach to better understand the dynamics of cold
collisionless fluids. We apply it by post-processing cosmological N--body
simulations and document the significant improvement it represents
over previous attempts to quantify the macroscopic and microscopic
aspects of the dark matter fluid flow. In particular, we show new
results for density estimates, a dark matter ``entropy'', bulk
velocities, velocity distribution functions -- many of which are
computed here for the first time.  We are confident that our approach
to tracing the dark matter sheet in phase-space gives important
physical insights which were inaccessible with previous approaches.

\section*{Acknowledgements}
T.A. is grateful for numerous conversations with Greg
  Bryan in the past ten years on how one might solve for dark matter dynamics
  directly in phase-space and acknowledges support by the National
  Science foundation through award number AST-0808398 and the LDRD
  program at the SLAC National accelerator laboratory as well as the
  Terman fellowship at Stanford University. He also acknowledges help
  by Patrick Abel in constructing tetrahedra from paper which helped
  considerably in understanding the many possible options of
  tessellations of the dark matter sheet.

\bibliography{./MyReferences} 
\label{lastpage}

\end{document}